\pdfpageattr {/Group << /S /Transparency /I true /CS /DeviceRGB>>}  
\documentclass[aps,pra,10pt,reprint,showpacs]{revtex4-1}

\usepackage[latin1]{inputenc}

\usepackage[x11names]{xcolor}
\usepackage{amsmath}
\usepackage{amssymb}
\usepackage{graphicx}
\usepackage{dsfont}
\usepackage{multirow}
\usepackage[export]{adjustbox}
\usepackage{ulem}
\usepackage{subfigure}

\usepackage{placeins}

\usepackage{marvosym}

\newcommand{\1}{\mathds{1}}

\newcommand{\ket}[1]{| #1 \rangle}
\newcommand{\bra}[1]{\langle #1 |}

\DeclareMathOperator{\tr}{tr}

\newcommand{\Podd}{P_{\mathrm{odd}}}
\newcommand{\Poddtilde}{\tilde{P}_{\mathrm{odd}}}
\renewcommand{\emph}{\textit}

\PassOptionsToPackage{cmyk}{xcolor}

\begin{document}
\title{Large-scale quantum networks based on graphs}
\author{Michael Epping}
\email{epping@hhu.de}
\author{Hermann Kampermann}
\author{Dagmar Bru\ss}
\affiliation{Institut f\"{u}r Theoretische Physik III, Heinrich-Heine-Universit\"{a}t D\"{u}sseldorf, Universit\"{a}tsstr. 1, D-40225
D\"{u}sseldorf, Germany}

\begin{abstract}
Society  relies and depends increasingly on information exchange and communication. In the quantum world, 
security and privacy is a built-in feature for information processing. 
 The essential ingredient for exploiting these  quantum advantages is the resource of entanglement, which can be  shared between two or 
more parties. The distribution of entanglement over large distances  constitutes a key challenge for current research and development.
Due to   losses of the transmitted quantum particles,  which typically scale exponentially with the
distance, intermediate quantum repeater
 stations are needed.  
Here we show how to generalise the quantum repeater  concept to the multipartite case, by
fully describing large-scale quantum networks, i.e. network nodes and their long-distance links,
 in the language of graphs and graph states. This unifying  approach comprises both the distribution of multipartite entanglement across
the network, and the  protection against errors via encoding. 
The correspondence to graph states also provides a tool for optimising the architecture of quantum networks.
\end{abstract}
\maketitle
\noindent
\section{Introduction}
Quantum entanglement is one of the pillars of quantum information processing.
Distribution of entanglement among two or more spatially separated parties is a necessary ingredient for many tasks in quantum information theory, including distributed quantum
computing~\cite{Buhrman03}, blind quantum computing~\cite{Broadbent09}, teleportation~\cite{BBC93}, telecloning~\cite{Murao99}, secret sharing~\cite{Markham08} and quantum cryptography schemes~\cite{BB84,Chen04,Duer05}. Multipartite entanglement enables a violation of Bell inequalities that grows exponentially with the number of parties~\cite{Mermin1990}.
However, the  controlled distribution of entanglement, in particular of multipartite entanglement,
over long distances is a major challenge, due to unavoidable imperfections such as particle losses and decoherence.

The seminal idea of  quantum repeaters \cite{Briegel98,Duer99} is based on the
distribution of  short-range entanglement between intermediate repeater stations
 (thus avoiding losses that grow typically 
exponentially with the distance) and subsequent entanglement swapping, which
connects the short links along a line  to long-range bipartite entanglement. 
Several theoretical variations have
been proposed: some of them are based on entanglement distillation~\cite{Duan01,vanLoock06,Zwerger12} and others are based on forward error correction~\cite{Knill96,Jiang09,Fowler10,Muralidharan14}. Much experimental progress towards the realisation of
a quantum repeater has been made \cite{Cory98,Yuan08,Kimble08,Sangouard11,Schindler11,Clausen12,Hensen15}.

``Partially quantum'' networks are considered in the so-called trusted node scenario~\cite{SECOQC09}, while fully quantum networks have been investigated in the context of network routing~\cite{Elliott02,Acin07,VanMeter13,Perseguers13} and coding~\cite{Leung06,Hayashi07} strategies and heterogeneous network technologies~\cite{Nagayama15}.

\begin{figure*}[tbp]%
\centering%
\includegraphics[width=\linewidth]{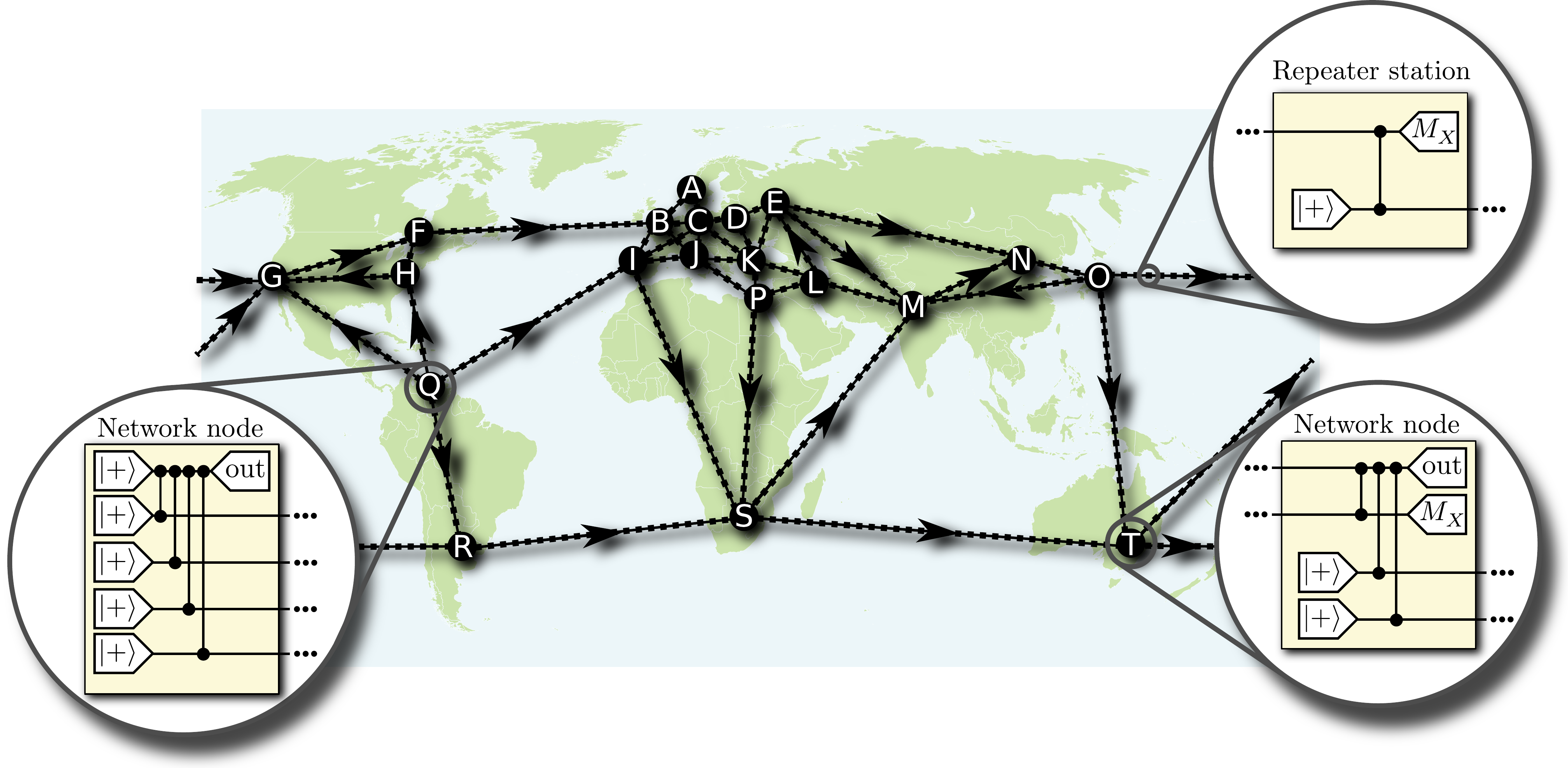}%
\caption{
Multipartite quantum network based on graphs: network nodes together with links
between them constitute a graph.  Both network nodes and repeater stations receive and send
quantum particles. They prepare qubits (in the $\ket{+}$-state), perform entangling quantum gates ($C_Z$-gates) and  measurements (in the $X$-basis). The number of such actions
for a given node depends on its number of neighbours. Arrows indicate the transmission direction. Some examples are illustrated.
Note that  repeater stations  have exactly two neighbours, while network nodes may have more than 
two neighbours.}\label{fig:weltkarte}
\end{figure*}

Here we propose a general multipartite quantum network architecture, 
where the long-distance links are bridged by quantum repeater stations.
This idea is illustrated in Fig.~\ref{fig:weltkarte} for the long-term vision of a ``world-wide quantum web''.
This network contains nodes (labelled by letters), which receive, measure and send particles. They could be located at, e.g., key institutions. Network nodes are connected by long-distance transmission channels, which are subdivided into shorter
channels by an appropriate number of quantum repeater stations - an example is also
shown in Fig.~\ref{fig:weltkarte}. 

Any network such as in Fig.~\ref{fig:weltkarte} forms a mathematical graph by identifying the network nodes with vertices and the quantum channels with edges. To any graph a corresponding graph state can be associated~\cite{Schlingemann01,Hein04}. These states
are highly entangled and constitute a valuable resource, e.g. for one-way quantum computation~\cite{Schlingemann01,Raussendorf03,Duer03,Hein04,Guehne05}.
In our proposal  the goal is to establish a multipartite entangled graph state
between the network nodes. This goal is reached in two steps. Step 1:  a graph state is produced,
where \emph{both} the network nodes and the repeater stations constitute vertices. Step 2:  the vertices
corresponding to the
repeater stations are ``removed'' by appropriate measurements. 
It is important to note that no memories are needed at the repeater stations, as the measurements can be performed immediately, as will be explained in more detail below.

In order to deal with unavoidable errors, for example photon losses in fibres or
in the atmosphere, quantum error correction will be used, i.e. the nodes and repeater stations will process higher-dimensional encodings of physical qubits;  we will use
stabiliser codes throughout this paper.

As the same language of stabilisers is used for both the encoding and the target states, our scheme of a global quantum repeater network is concise and general.

\section{From graphs to quantum repeater networks}
A  mathematical graph $G$ consists  of a set $V$ of $N$ vertices and a set E of edges, each of  which connects
two vertices, i.e. $E\subset
V\times V$. In Fig.~\ref{fig:weltkarte}, the network nodes as well as the repeater stations are  vertices, and
all transmission channels between them are  edges of a graph.  To each mathematical graph G 
  corresponds a   graph state $\ket{G}$, which
 can be defined in two equivalent ways. First, 
a graph state can be physically produced by switching on a specific entangling gate
 for each edge of the graph. Concretely, 
$\ket{G}$ is the state that is created from the state $\ket{+}^{\otimes N}$,
with 
$\ket{+}=\frac{1}{\sqrt{2}}(\ket{0}+\ket{1})$, by applying a controlled-phase gate $C_Z$ to each pair
$(i,j)$ of vertices in $E$, i.e. 
\begin{equation}
 \ket{G}=\prod_{(i,j)\in E} C_{Z}^{(i,j)} \ket{+}^{\otimes N}, \label{eq:ketG}
\end{equation}
where in the computational basis $\{\ket{00},\ket{01},\ket{10},\ket{11}\}$ the entangling gate reads
$C_Z=diag(1,1,1,-1)$.

Second, a graph state is the unique state which is eigenstate of a set of so-called stabiliser
operators, with eigenvalues +1. Each vertex $i$ of the graph has an associated stabiliser operator
$g_i$ which is a product of the Pauli-$X$ operator for vertex $i$ and   the Pauli-$Z$ operator for 
all its neighbours, i.e. $g_i$  reads 
\begin{equation}
 g_i=X_i \prod_{\substack{j\in V\\(i,j)\in E}} Z_j.\label{eq:gi}
\end{equation}
Here, $X_i$ is a short-hand notation for the Pauli operator $X$ acting on vertex $i$ and the identity $\1$ at all other vertices.
The graph state $\ket{G}$ is defined via the eigenequations  $g_i \ket{G} = \ket{G}$, for all $i\in V$. 
Note that a product of stabiliser operators is also a stabiliser.

In order to present our main idea, let us first describe the two mentioned steps. In
step 1, a graph state according to the graph in Fig.~\ref{fig:weltkarte} is created: for the simple line graphs, which
constitute the long-distance links, each repeater station receives one qubit from the previous station,
produces one qubit in  state $\ket{+}$,  entangles it with the  qubit from the previous station via 
  a $C_Z$ gate and then sends  the second qubit through the channel to the next repeater station, which
acts in the same way. Thus, the edges between repeater stations are created. The network nodes act in a slightly
different way: depending on their number of neighbours,
 they receive a certain number of inputs, create a certain number of qubits in state $\ket{+}$,
perform entangling $C_Z$ gates, and send on the appropriate number of qubits to the neighbouring repeater 
stations. Some examples are given in Fig.~\ref{fig:weltkarte}. Thus, the whole graph of Fig.~\ref{fig:weltkarte} will be produced.

In step 2, the vertices corresponding to all  repeater stations are removed by a simple Pauli $X$-measurement at each 
repeater station. The reasoning is as follows: remember that a product of stabilisers is
also a stabiliser.
Consider the vertex S (South Africa) in the quantum network shown in Fig.~\ref{fig:weltkarte} and assume that the number of repeater stations is even on each edge (odd numbers can be treated in an analogous way). Take the product of the stabiliser generators starting from S and for every second repeater station, until reaching the neighbours T, M, P, I, and R. Due to the definition of $g_i$ in equation~(\ref{eq:gi}) and the fact that $Z^2=\1$, this product of stabilisers contains only $X$-operators at S and the mentioned repeater stations, and $Z$-operators on the neighbouring network nodes of S in the network (see also \cite{Wu15}). We call this stabiliser operator the main stabiliser centred on S. Measuring all repeater stations in the $X$-basis projects the state onto one stabilised by $\pm X_S Z_T Z_M Z_P Z_I Z_R$ in the Hilbert space of the network nodes only. Here the sign of the stabilizer operator depends on the parity of the measurement outcomes of the repeater stations included in the main stabiliser centred on S. The minus sign can be removed by applying the so-called by-product operator $Z_S$ in this case. This reasoning holds for all network nodes. By comparison of the obtained stabilisers with equation~(\ref{eq:gi}) it is clear that the graph state corresponding to the global network (large vertices in Fig.~\ref{fig:weltkarte}) has been produced.

Even though we have explained the procedure in two consecutive steps, it is not necessary to
store the full graph state: as the local measurements commute with the operations on other 
repeater stations, a qubit can be measured immediately after action of the $C_Z$ gate, which
is easier to realise experimentally. Thus, 
the whole graph state between network nodes is gradually built up in a one-way
fashion, as explained in Fig.~\ref{fig:weltkarte}, without
need for memories in the repeater stations.

In an implementation of the above scheme, losses in the transmission channels, noise
in the gates as well as errors in preparation and measurement will occur and would lead
to a low-quality output state. As a solution to this problem, quantum  error correction can be employed:
The main idea is to encode the state of a so-called logical qubit redundantly into many physical qubits,
such that a local error leads to a unique error syndrome and can be corrected
by applying a suitable operation~\cite{LidarBrunQEC13}. This is in contrast to previous ideas~\cite{Zwerger14} where
graph states were used as resource states for measurement-based
implementations of quantum error correction. 

\begin{figure}[tbp]%
\centering%
\includegraphics[width=\linewidth]{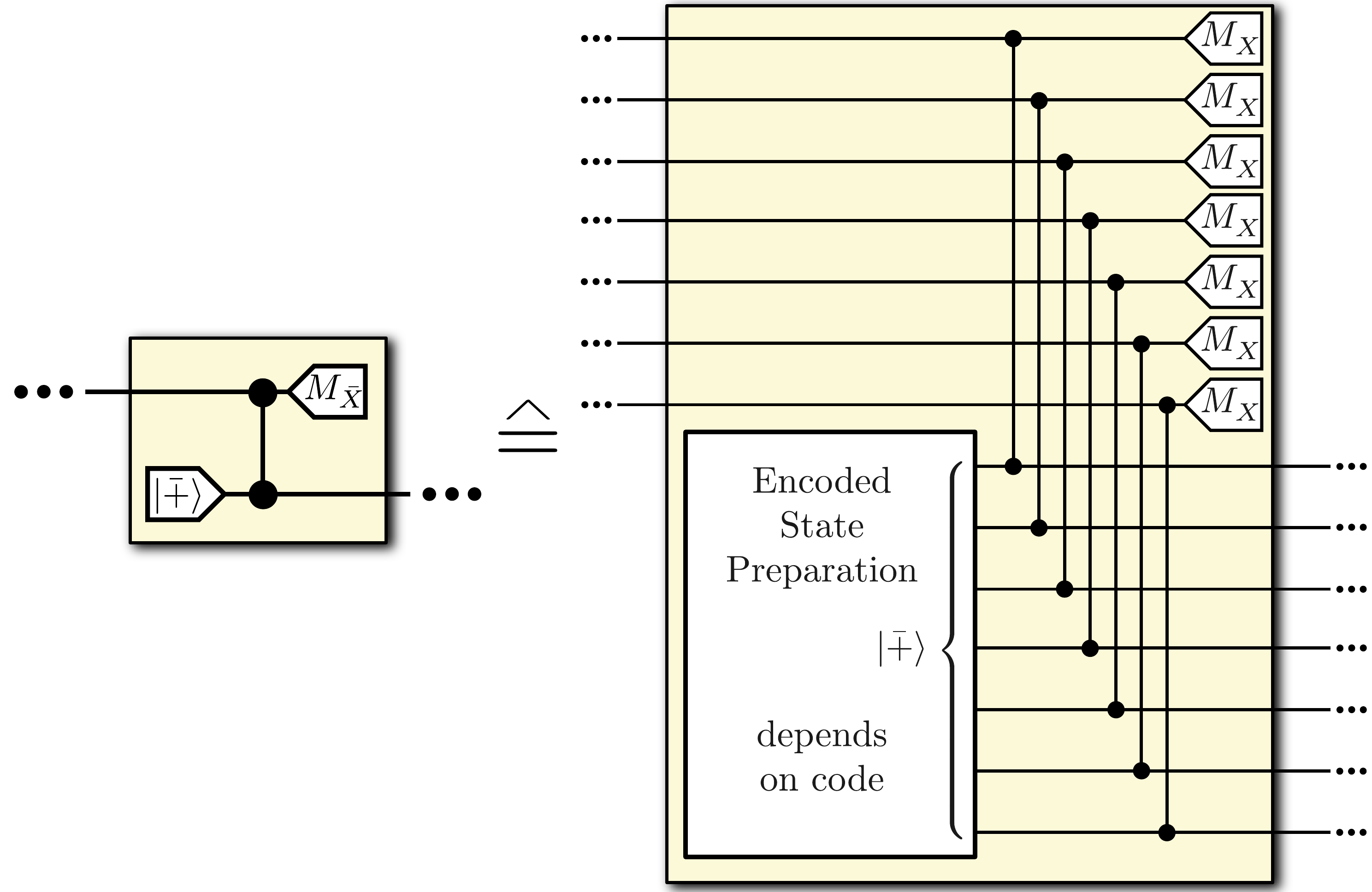}%
\caption{
Repeater station with encoding: encoded state preparation, action of gates and 
measurements on physical qubits correspond to the equivalent short-hand notation of  actions
on logical qubits, denoted by a bar. 
}\label{fig:logicalrepeater}
\end{figure}

In the present article we make use of so-called stabiliser codes~\cite{Gottesman97}, which are defined via a set of stabiliser operators,
the eigenstates of which (with eigenvalue 1) are the codewords. In particular, we focus here on a subclass of stabiliser codes, the Calderbank-Shor-Steane (CSS) codes~\cite{LidarBrunQEC13} which have the useful property of ``transversal'' logical entangling gates (see Appendix~\ref{sec:errprop}). Thus the only change in our
scheme is that instead of
 initial physical qubit states $\ket{+}$  multi-qubit encoded logical states, denoted in the following as
 $\ket{\bar +}$, need to be generated. The specific structure of $\ket{\bar +}$  depends on the chosen
 error correction code. Low-error state preparation can be done more efficiently than general quantum operations on an unknown state, see e.g. \cite{Paetznick11} for a preparation scheme for the quantum Golay code~\cite{Golay49,Goethals71,Elia95}. 
The  repeater operation in the encoded case and the short-hand notation $\ket{\bar +}$
and $M_{\bar X}$  for the measurement on the encoded state is illustrated in Fig.~\ref{fig:logicalrepeater}.

\section{Error analysis in the graph language}

The unified description of the quantum network in terms of stabilisers for both the states and 
the encoding allows for a comprehensive analysis of errors and performance study.
An error can be noticed, in the sense that it is known which qubit is affected (e.g. a no-detection event), or unnoticed (noise).
For our performance study we will use the usual exponential loss model in optical
fibres, i.e. the failure probability during transmission $f_T$ is given by
\begin{equation}
 f_{T}=1-(1-f_{C}) e^{-L_0/L_{\textrm{att}}},
\end{equation}
where $f_{C}$ is a coupling failure probability, $L_0$ is the distance between
repeater stations and $L_{\textrm{att}}$ is  the attenuation length of the fibre,
for which we
will use the value  $L_{\textrm{att}}= 20\;\textrm{km}$.
All qubit errors (for sources, gates, channels, detectors) will be modelled by the depolarizing channel,
characterised as
\begin{equation}
 \rho \rightarrow (1-f) \rho + f\frac{1}{2}\1,
\end{equation}
 i.e. with a failure
probability $f$ the state of the qubit is proportional to the identity, and with probability $(1-f)$ it
is unaffected. Thus in case of failure the state is depolarized to the completely mixed state. The same effect is achieved by randomly applying bit-flips and phase-flips to the state. This mathematically equivalent viewpoint of a perfect operation followed by discrete $X$ and $Z$ errors is very convenient~\cite{nielsen00}.

When a  physical $X$ or $Z$ error has occurred, it  propagates via the gates through the network,
 i.e. it may influence
the consecutive qubits and measurement results. However, fortunately the spreading of errors  along a repeater line
is restricted to a finite
length, concretely to up to two repeater stations. This is due to   some simple rules
for error propagation  via a $C_Z$ gate:  a $Z$ error that occurs on one of the two input
qubits of a $C_Z$ gate remains a $Z$ error on the corresponding output qubit and does not affect the other output qubit.
An $X$ error that occurs on one of the two input
qubits remains an $X$ error on the corresponding output qubit, but also causes a $Z$ error  on the other output qubit.
 (If more than one error has occurred, the output qubits will suffer from corresponding products
of errors.)
An $X$ or $Z$ error may therefore be spread to the next repeater station, where the 
corresponding qubit will pass through the next $C_Z$ gate and then will be measured.
Regarding the measurement, an $X$- ($Z$-)error before an  $X$- ($Z$-)measurement does not
affect the measurement outcome, while an $X$- ($Z$-)error before a  $Z$- ($X$-)measurement
flips the measurement outcome. 

\begin{figure}[tbp]%
\centering%
\includegraphics[width=\linewidth]{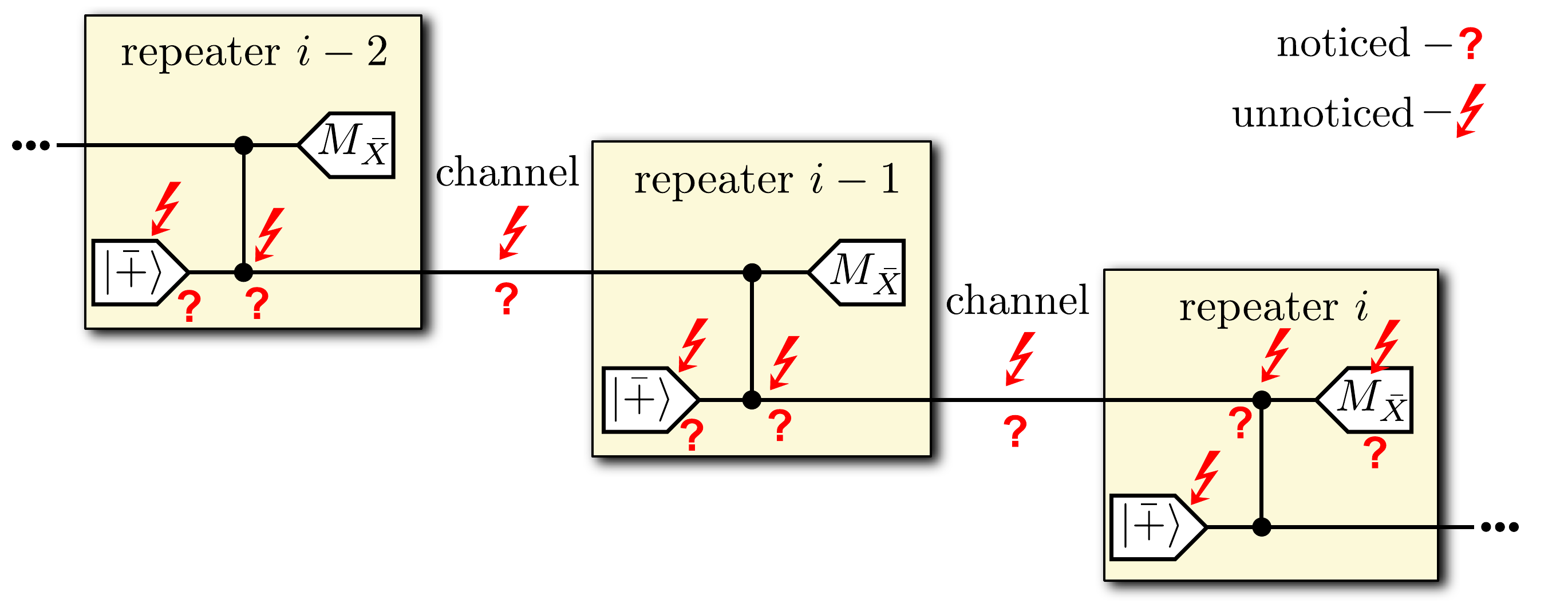}%
\caption{
Error propagation through a graph state repeater line: 
All possible sources of an error at repeater station $i$ are shown. Here,
 '?'  denotes noticed errors and '\Lightning' denotes unnoticed errors. The measurement outcome at repeater station
 $i$  is marked as '?' in case of a noticed error, and may be flipped in case
of an unnoticed error.}\label{fig:circuit}
\end{figure}

The possible sources of errors are shown in Fig~\ref{fig:circuit}. It is important to note, due to the arguments
given above, that repeater station number $i$ is only influenced by errors propagating
from nearest and next-to-nearest neighbours, i.e. from stations $(i-1)$ and $(i-2)$. 

If an error is noticed, the corresponding measurement outcome is set to ``$?$''. The physical error rate depends on the failure probabilities for transmission,
gates and measurements and is explicitly calculated in Appendix~\ref{sec:errprop}.

In Fig.~\ref{fig:circuit} we focus on the physical error rates along a repeater line. The generalisation of our analysis to more gates and more qubits in the case of the network nodes is straightforward and can be described in terms of the vertex (in- and out-)degree, see Appendix~\ref{sec:generalization}. Note, however, that in a large-scale quantum network there are many more repeater stations than network nodes. Thus the performance of the network mainly depends on the error rate at the repeater stations, which was described above.

Up to now we have described the physical errors. For a given encoding the logical error rate, i.e. the rate of uncorrectable errors, is a function of the physical error rate, see Appendix~\ref{sec:fqbar}.

Remember that the (logical) measurement outcomes at the repeater stations of the main stabiliser centred on a node $v$ determine whether the by-product operator $Z_v$ needs to be applied. Thus even numbers of logical errors on the corresponding repeater stations cancel each other. The local error rate $e_v$ at the vertex $v$ is an important indicator of the quality of the produced state. The formula to calculate $e_v$ follows the previous reasoning and is given in Appendix~\ref{sec:logicalgraphstates}. The error rates $e_v$ effectively combine all errors of the repeater stations and simplify the analysis considerably. The stabiliser error rates allow to bound the fidelity of the established state, see Appendix~\ref{sec:logicalgraphstates}.

\section{Performance and quantum network architectures}
The performance of a quantum network may depend on the task it was built for: possible  figures of merit are e.g. the rate
for the production of a long-distance entangled state, the success probability for a task such as quantum teleportation,
or the secret key rate in a cryptographic setting. In the following we will use as figure of merit the cost-performance ratio $C$
which compares the needed resources~\cite{Muralidharan14}: 
It is defined as the total number of needed qubits divided by
the total distance $L$ and a specific quality factor $Q$, i.e. 
\begin{equation}
C=\frac{n w}{L Q},\label{eq:costs}
\end{equation}
where $n$ is the number of qubits per station (neglecting preparation overhead) and $w$ is the number of repeater stations.
\begin{figure*}[htbp]
\centering
\subfigure[The network with a cycle on the left produces the same final state as the network without cycle on the right (up to local basis transformations).]{\includegraphics[scale=0.5]{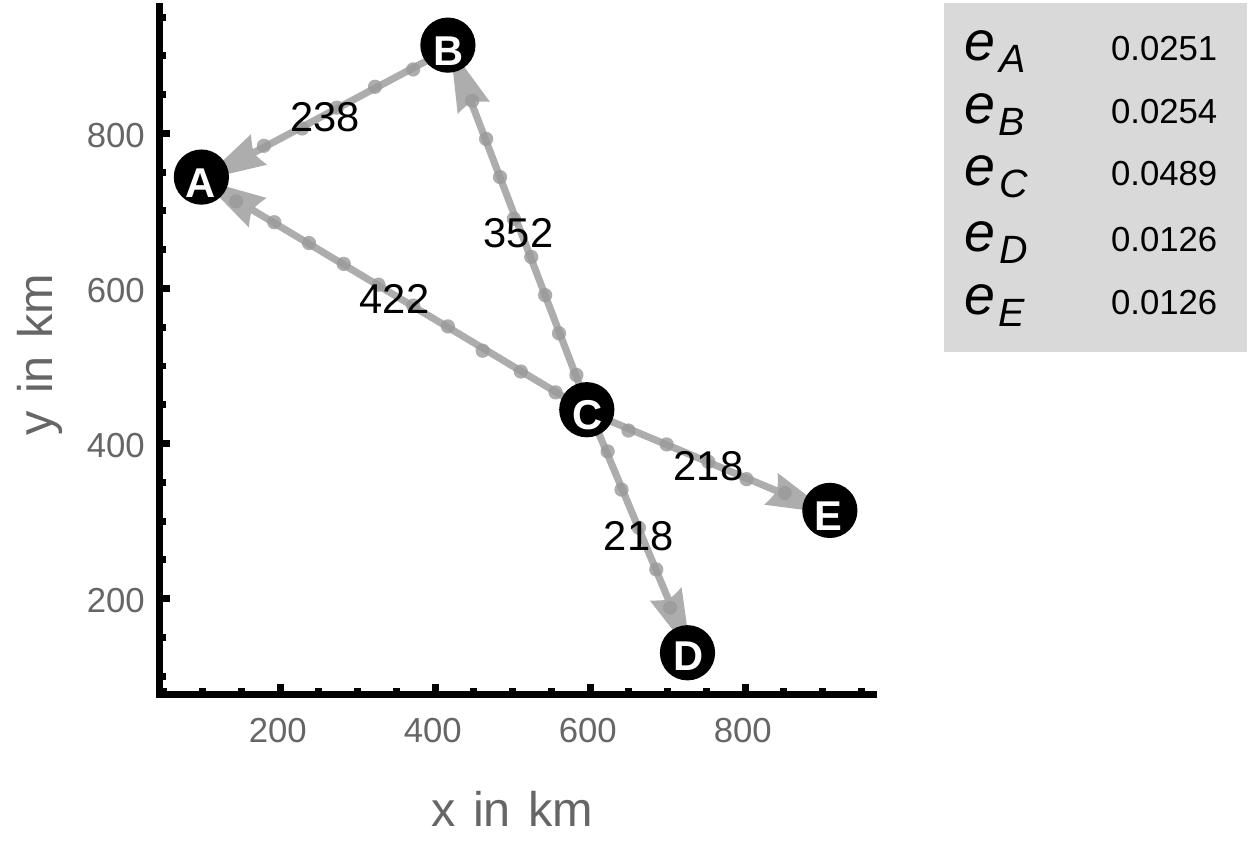}\qquad \includegraphics[scale=0.5]{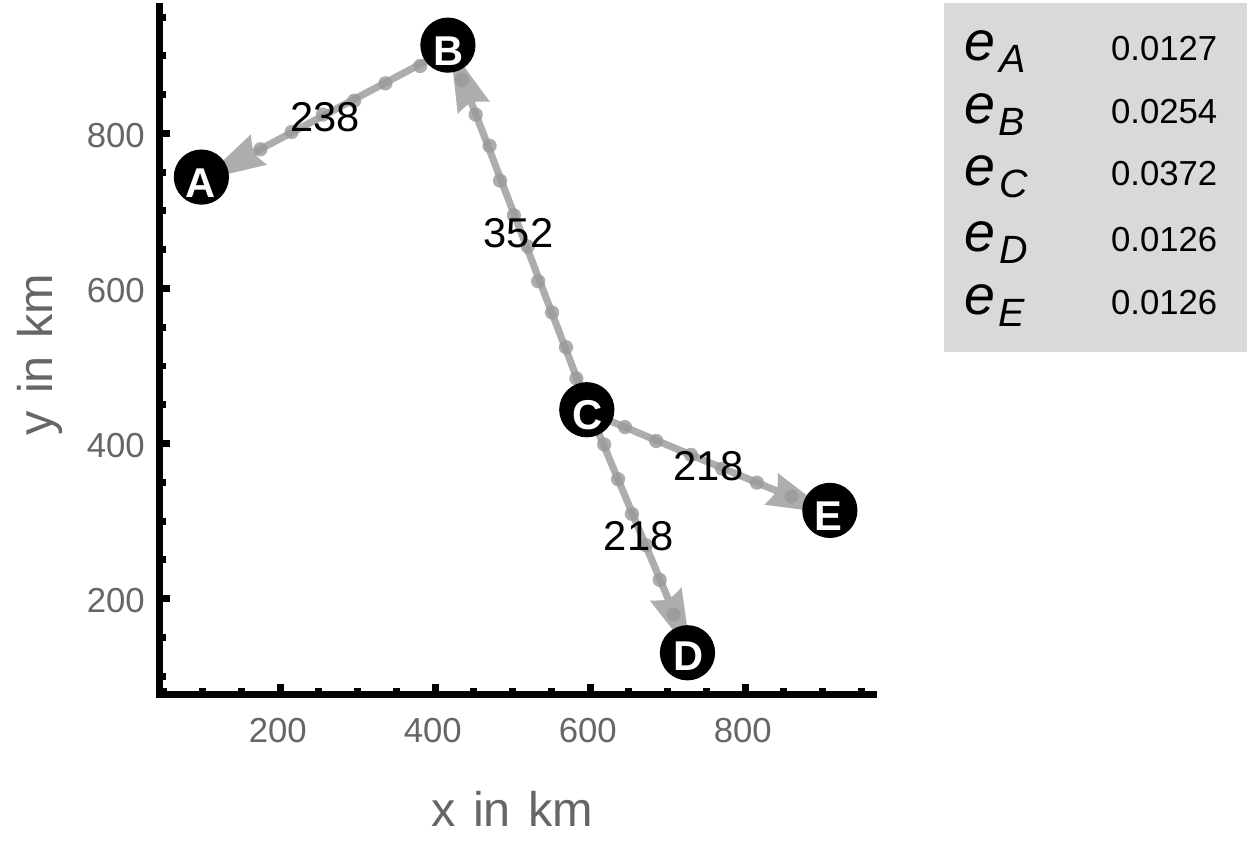}\label{fig:gA}}\\
\subfigure[While both networks have the same number of links and show comparable performance (in terms of the error rates of the final state), the right requires less repeater stations.]{\includegraphics[scale=0.5]{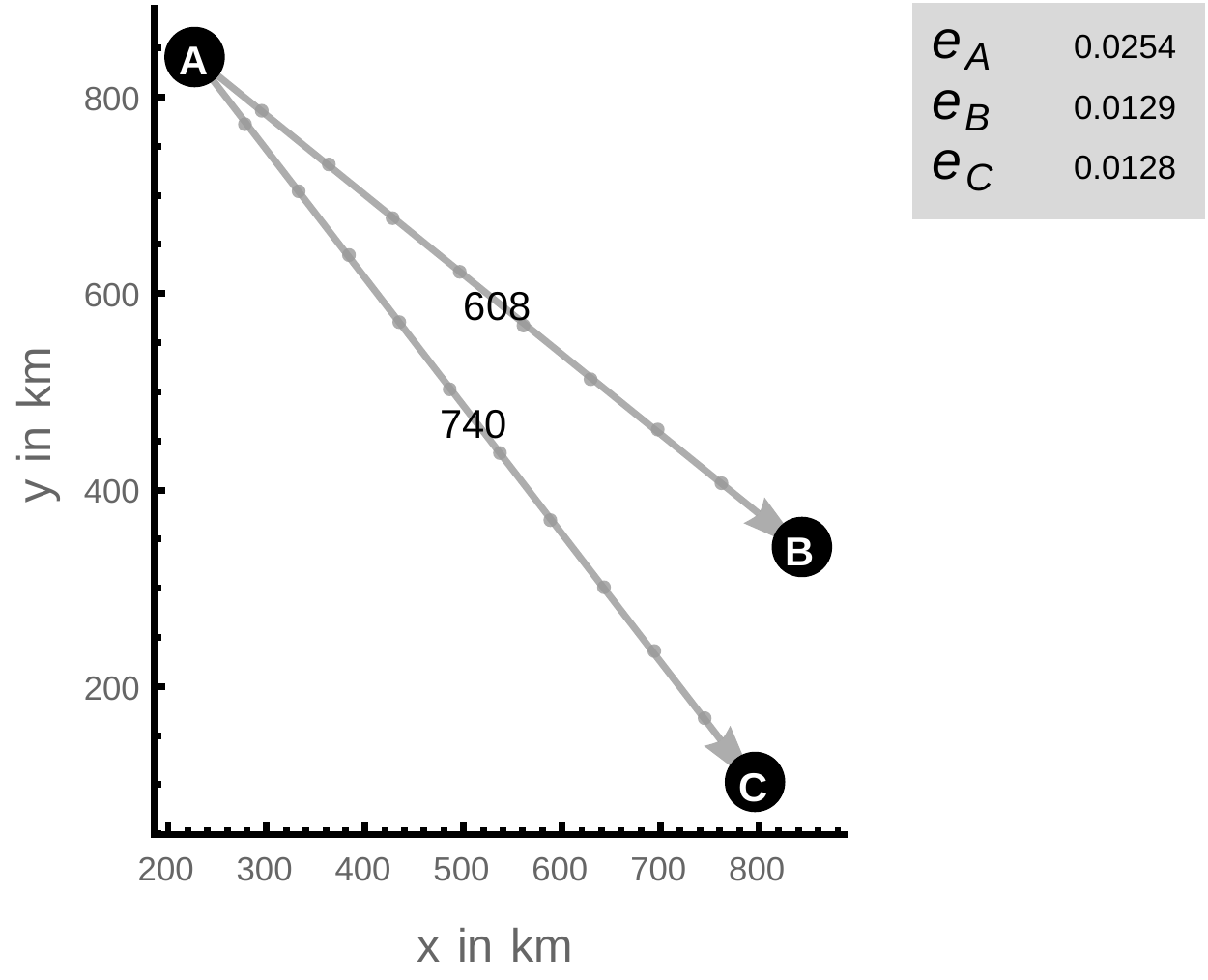}\qquad \includegraphics[scale=0.5]{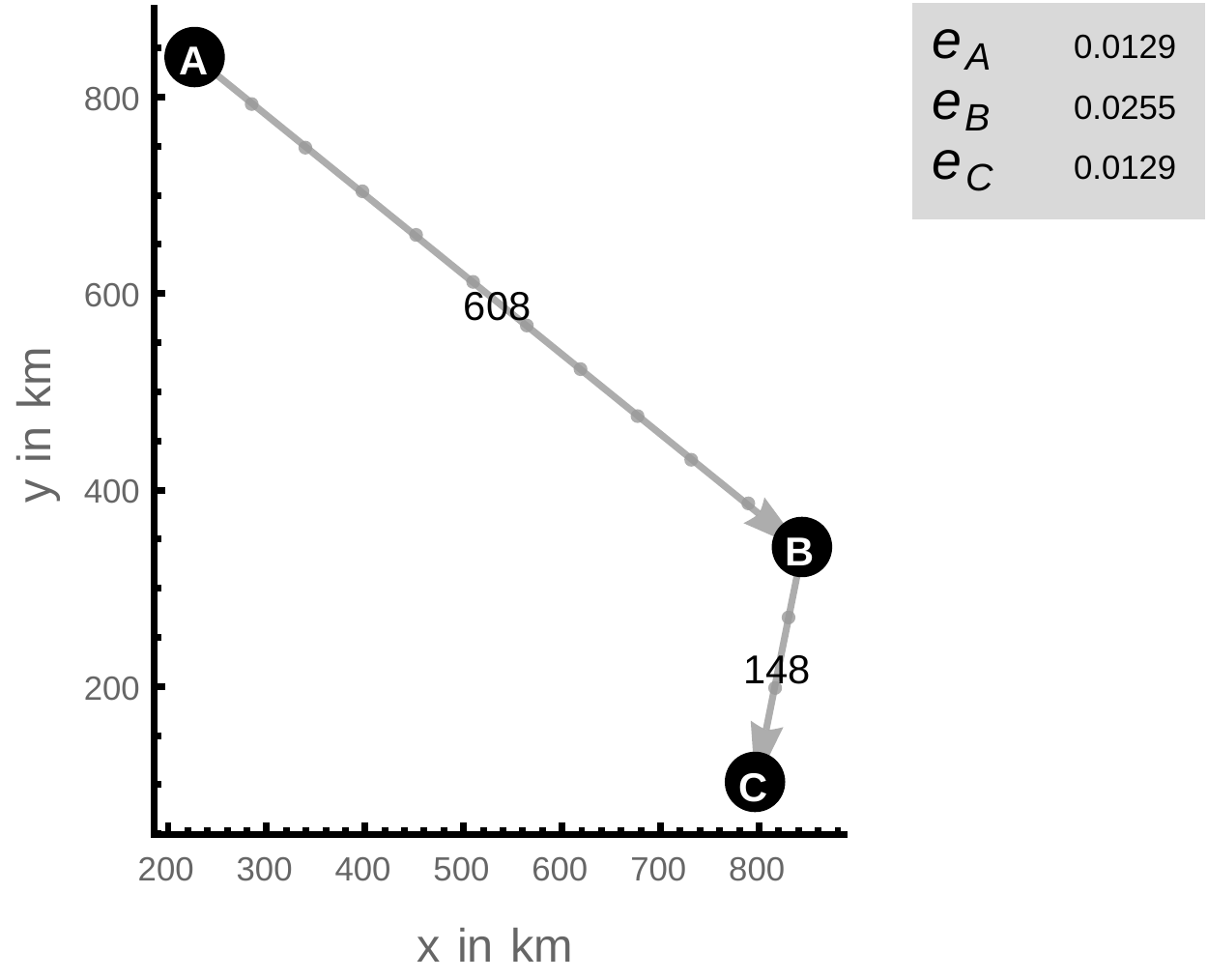}\label{fig:gC}}\\
\subfigure[The left network requires less repeaters, because the total length of the repeater lines is shorter. The right network, however, has a better performance in terms of the error rates, because the maximal vertex degree is lower.]{\includegraphics[scale=0.5]{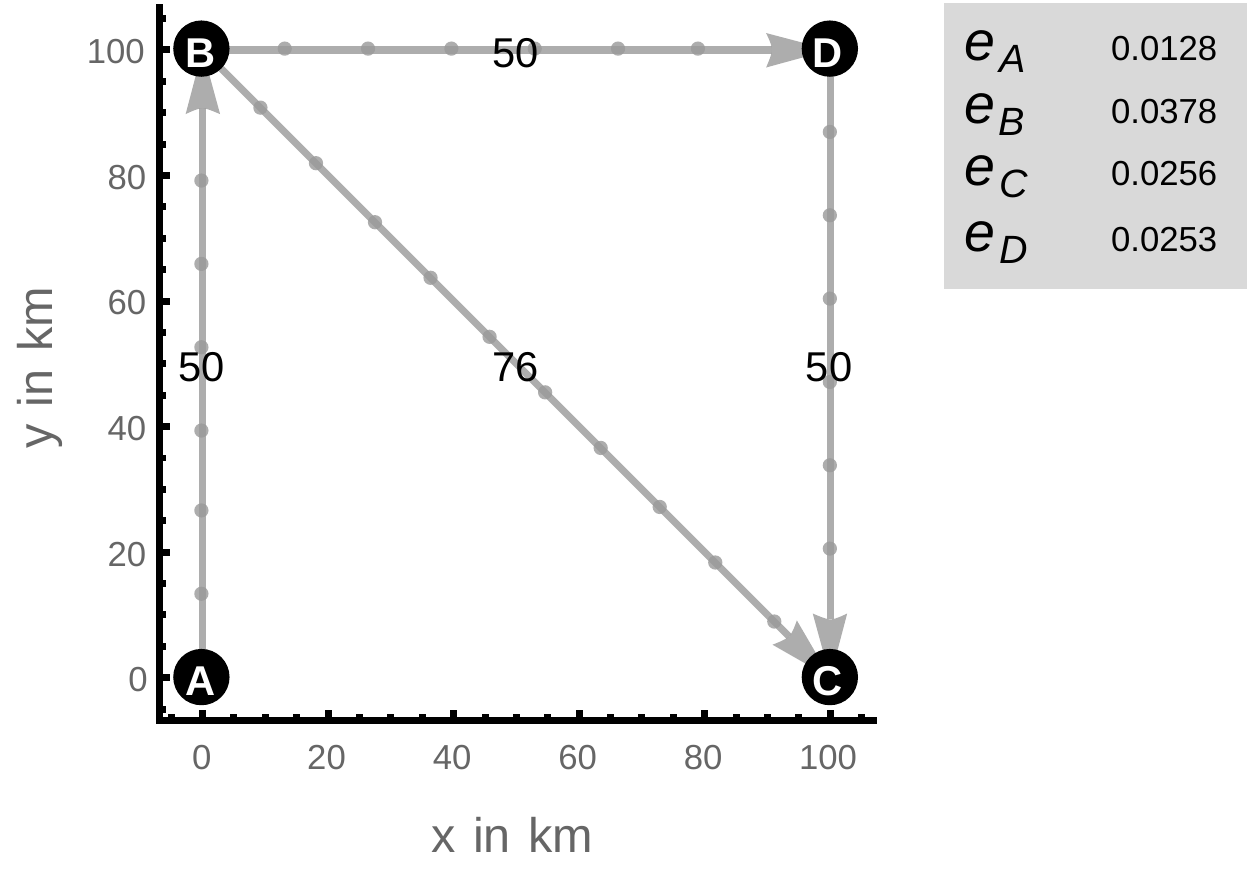}\qquad \includegraphics[scale=0.5]{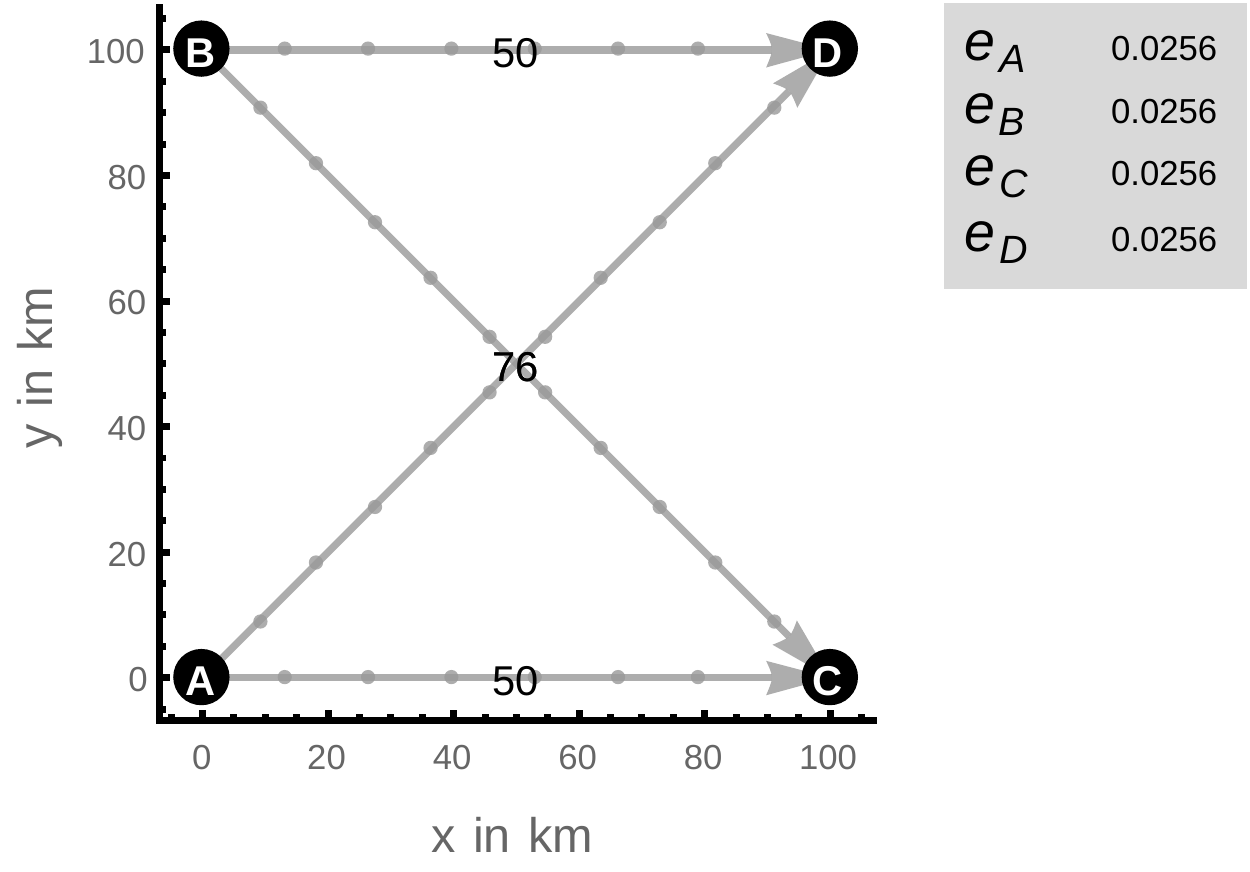}\label{fig:gD}}\\
\caption{The left example networks are local-unitary-equivalent to their right hand side counterparts. Despite producing the same final state up to local basis transformations, these networks can differ in their performance and cost. For all figures a gate error rate of $f_G=1\times 10^{-4}$ has been used. The number of repeater stations on each link minimizes its cost-performance ratio $C$ and is shown as the edge weight. The resulting error rates at the network nodes are shown in the box next to the network. They are approximately proportional to the corresponding vertex degree, because all links introduce roughly the same amount of errors (as a result of the optimization w.r.t. the cost-performance ratio $C$).}\label{fig:LCGraphen}
\end{figure*}
Our description in the graph state language provides a tool  to optimise the architecture
of a quantum  network: two graphs $G$ and $G'$ with the same set of vertices $V$ but different sets of edges $E\neq E'$
may correspond to local-unitary equivalent graph states~\cite{Hein04}, i.e. states that are related by local basis changes.
This fact leads to  general  optimisation arguments for quantum networks;  we now  consider only network nodes
as vertices and their connecting repeater lines as edges, which have a weight according to the number of repeater stations on this line.
\begin{enumerate}
 \item The graph $G'$ can have fewer edges than $G$.  
This corresponds to a reduced number of repeater lines in a network, see e.g. Fig.~\ref{fig:gA}. 
 \item The graph $G'$ can have less cycles than $G$, see e.g. Fig.~\ref{fig:gA} and \ref{fig:gD}. Note that in general cycles increase the required coherence time of the used quantum memories at the network nodes, because the qubits need to be stored until all gates have been applied to them.
 \item The total length of the edges in $E$ and $E'$ may differ, even if the number of edges of $G'$ and $G$ are equal,   see e.g.
Fig.~\ref{fig:gC}. One can thus  minimize the total number of repeater stations in the network.
 \item The maximal vertex degree of $G$ and $G'$ may be different. Given that the number of repeater stations is optimized w.r.t. $C$ for each individual edge, the rate of errors increases with the vertex degree (see Appendix~\ref{sec:logicalgraphstates}). Thus it can be advantageous to reduce the maximal vertex degree, see e.g. Fig.~\ref{fig:gD}.
\end{enumerate}
In order to illustrate the general performance of a graph state quantum repeater and to
compare different codes we consider in the following a bipartite setting, i.e. one repeater line.
This is a typical quantum cryptographic scenario, and therefore we use as the quality factor $Q$ in the
cost-performance ratio $C$ the effective secret fraction  $R$, given by
\begin{equation}
R=P_{\textrm{succ}} r_{\infty} \label{eq:R}
\end{equation}
where  $P_{\textrm{succ}}$ denotes the probability for not aborting of the protocol (one might choose to abort in case of a ``fatal'' pattern of noticed errors in order to increase the quality of the produced state),
employing a given code.  The factor $r_{\infty}$ is the secret fraction for the BB84 protocol, given by~\cite{Scarani09}
\begin{equation}
r_{\infty}=\max\{1-h(e_A)-h(e_B),0\}. \label{eq:rinfty}
\end{equation}
Here,  $e_A$ and $e_B$ are the 
error rates of the two end nodes, and  the binary entropy is  defined as $h(p)=-p \log_2(p) -(1-p) \log_2(1-p)$. Note that $e_A$ and $e_B$ depend on the logical error rate of the error correction code: a logical measurement error remains, if the outcomes are decoded to a codeword with wrong parity.\\

\begin{figure}[htbp] %
\centering %
\includegraphics[width=1.0\linewidth]{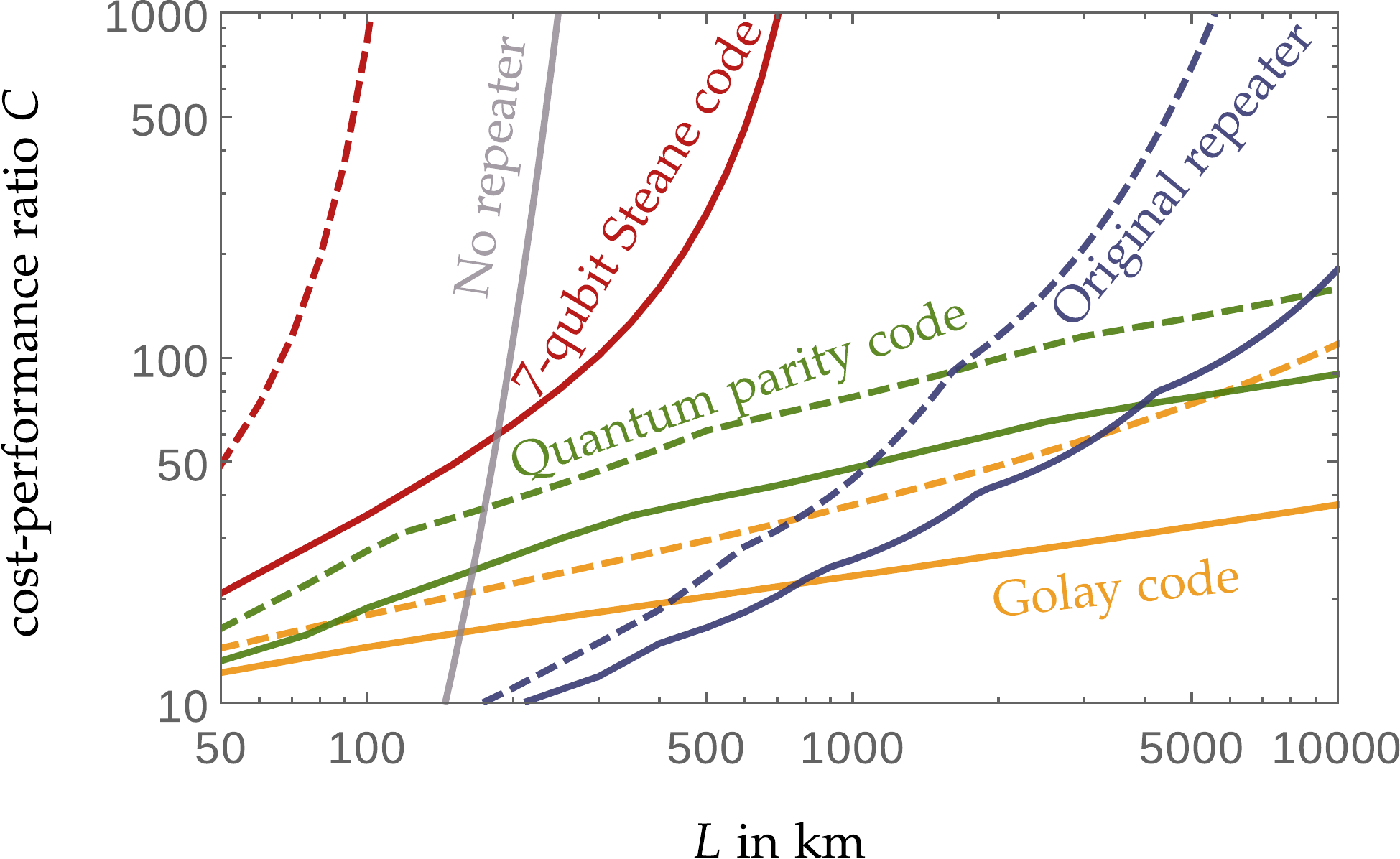} %
\caption[Cost of several codes]{
Comparison of the cost-performance ratio $C$ for  several codes: Encoding via the Seven-Qubit Steane code (red), the 23-qubit quantum Golay code (yellow),
and the (here up to 84-qubit) quantum parity code of \cite{Muralidharan14} (green), compared with the original (distillation based) scheme~\cite{Briegel98,Abruzzo13} (blue).
 The gate failure rates are $f_G=10^{-3}$ (dashed) and $f_G=10^{-4}$ (solid).
 The grey %
line corresponds to using no repeaters. The comparison with the original scheme assumes measurement times of $10\;\mathrm{\mu s}$.
} %
\label{fig:costs} %
\end{figure}

We optimized the cost-performance ratio $C$ with respect to the number of repeater stations
  $w$ for different $L$ for several codes. The optimal separation distance
of the repeater stations decreases with increasing total distance.
Note that the number of repeater stations in Fig.~\ref{fig:LCGraphen}, i.e. the optimal weight of the long-distance
edges, was also calculated using this bipartite figure of merit for each edge. 
For a comparison of various encoding schemes with the original repeater 
see Figure~\ref{fig:costs}. 
For distances larger than about 800 km, the Golay code outperforms all previous approaches.
With our methods, this type of comparison can now be performed for
any quantum network architecture and any quantum information processing task, using a corresponding figure of merit for the quality factor $Q$.

\section{Discussion}
Establishing a large-scale entangled quantum state  is
a formidable future task. In our proposal of a  graph state quantum repeater network this task 
finds a unified description in the elegant  language of stabilisers. Though being of abstract 
mathematical origin, this approach allows  to  quantitatively evaluate and
compare different implementations of any quantum network. For given quantum hardware such
as sources, transmission channels, gates and detectors, a suitable error correction code
can be found, and the performance for quantum information processing protocols
such as e.g. secret key generation between two or more
parties can be determined. 

For fixed locations of participating parties, our method
helps to design an optimal quantum network in terms of resources and performance with respect to a specific task (e.g. cryptography or synchronisation of distributed clocks~\cite{Chuang00,Komar14}).
Exploiting local unitary equivalence of different quantum networks has no classical counterpart and deserves further detailed investigations.
Future research on quantum networks will benefit from the presented description in the stabiliser formalism. This includes in particular the analysis of the efficient use of the whole network infrastructure to produce entangled states shared by a subset of parties.
While we showed that the performance of the 23-qubit quantum Golay code is outstanding for large distances, further research may focus on smaller codes for smaller networks.

\FloatBarrier

\begin{acknowledgments}
M.E. acknowledges helpful discussions with S. Muralidharan and financial support by the German Federal Ministry of Education and Research (BMBF).
\end{acknowledgments}
\bibliographystyle{apsrev4-1}

\begin{thebibliography}{51}%
\makeatletter
\providecommand \@ifxundefined [1]{%
 \@ifx{#1\undefined}
}%
\providecommand \@ifnum [1]{%
 \ifnum #1\expandafter \@firstoftwo
 \else \expandafter \@secondoftwo
 \fi
}%
\providecommand \@ifx [1]{%
 \ifx #1\expandafter \@firstoftwo
 \else \expandafter \@secondoftwo
 \fi
}%
\providecommand \natexlab [1]{#1}%
\providecommand \enquote  [1]{``#1''}%
\providecommand \bibnamefont  [1]{#1}%
\providecommand \bibfnamefont [1]{#1}%
\providecommand \citenamefont [1]{#1}%
\providecommand \href@noop [0]{\@secondoftwo}%
\providecommand \href [0]{\begingroup \@sanitize@url \@href}%
\providecommand \@href[1]{\@@startlink{#1}\@@href}%
\providecommand \@@href[1]{\endgroup#1\@@endlink}%
\providecommand \@sanitize@url [0]{\catcode `\\12\catcode `\$12\catcode
  `\&12\catcode `\#12\catcode `\^12\catcode `\_12\catcode `\%12\relax}%
\providecommand \@@startlink[1]{}%
\providecommand \@@endlink[0]{}%
\providecommand \url  [0]{\begingroup\@sanitize@url \@url }%
\providecommand \@url [1]{\endgroup\@href {#1}{\urlprefix }}%
\providecommand \urlprefix  [0]{URL }%
\providecommand \Eprint [0]{\href }%
\providecommand \doibase [0]{http://dx.doi.org/}%
\providecommand \selectlanguage [0]{\@gobble}%
\providecommand \bibinfo  [0]{\@secondoftwo}%
\providecommand \bibfield  [0]{\@secondoftwo}%
\providecommand \translation [1]{[#1]}%
\providecommand \BibitemOpen [0]{}%
\providecommand \bibitemStop [0]{}%
\providecommand \bibitemNoStop [0]{.\EOS\space}%
\providecommand \EOS [0]{\spacefactor3000\relax}%
\providecommand \BibitemShut  [1]{\csname bibitem#1\endcsname}%
\let\auto@bib@innerbib\@empty
\bibitem [{\citenamefont {Buhrman}\ and\ \citenamefont
  {Röhrig}(2003)}]{Buhrman03}%
  \BibitemOpen
  \bibfield  {author} {\bibinfo {author} {\bibfnamefont {H.}~\bibnamefont
  {Buhrman}}\ and\ \bibinfo {author} {\bibfnamefont {H.}~\bibnamefont
  {Röhrig}},\ }in\ \href {\doibase 10.1007/978-3-540-45138-9_1} {\emph
  {\bibinfo {booktitle} {Mathematical Foundations of Computer Science 2003}}},\
  \bibinfo {series} {Lecture Notes in Computer Science}, Vol.\ \bibinfo
  {volume} {2747},\ \bibinfo {editor} {edited by\ \bibinfo {editor}
  {\bibfnamefont {B.}~\bibnamefont {Rovan}}\ and\ \bibinfo {editor}
  {\bibfnamefont {P.}~\bibnamefont {Vojtas}}}\ (\bibinfo  {publisher} {Springer
  Berlin Heidelberg},\ \bibinfo {year} {2003})\ pp.\ \bibinfo {pages}
  {1--20}\BibitemShut {NoStop}%
\bibitem [{\citenamefont {Broadbent}\ \emph {et~al.}(2009)\citenamefont
  {Broadbent}, \citenamefont {Fitzsimons},\ and\ \citenamefont
  {Kashefi}}]{Broadbent09}%
  \BibitemOpen
  \bibfield  {author} {\bibinfo {author} {\bibfnamefont {A.}~\bibnamefont
  {Broadbent}}, \bibinfo {author} {\bibfnamefont {J.}~\bibnamefont
  {Fitzsimons}}, \ and\ \bibinfo {author} {\bibfnamefont {E.}~\bibnamefont
  {Kashefi}},\ }in\ \href {\doibase 10.1109/FOCS.2009.36} {\emph {\bibinfo
  {booktitle} {Foundations of Computer Science, 2009. FOCS '09. 50th Annual
  IEEE Symposium on}}}\ (\bibinfo {year} {2009})\ pp.\ \bibinfo {pages}
  {517--526}\BibitemShut {NoStop}%
\bibitem [{\citenamefont {Bennett}\ \emph {et~al.}(1993)\citenamefont
  {Bennett}, \citenamefont {Brassard}, \citenamefont {Cr\'epeau}, \citenamefont
  {Jozsa}, \citenamefont {Peres},\ and\ \citenamefont {Wootters}}]{BBC93}%
  \BibitemOpen
  \bibfield  {author} {\bibinfo {author} {\bibfnamefont {C.~H.}\ \bibnamefont
  {Bennett}}, \bibinfo {author} {\bibfnamefont {G.}~\bibnamefont {Brassard}},
  \bibinfo {author} {\bibfnamefont {C.}~\bibnamefont {Cr\'epeau}}, \bibinfo
  {author} {\bibfnamefont {R.}~\bibnamefont {Jozsa}}, \bibinfo {author}
  {\bibfnamefont {A.}~\bibnamefont {Peres}}, \ and\ \bibinfo {author}
  {\bibfnamefont {W.~K.}\ \bibnamefont {Wootters}},\ }\href {\doibase
  10.1103/PhysRevLett.70.1895} {\bibfield  {journal} {\bibinfo  {journal}
  {Phys. Rev. Lett.}\ }\textbf {\bibinfo {volume} {70}},\ \bibinfo {pages}
  {1895} (\bibinfo {year} {1993})}\BibitemShut {NoStop}%
\bibitem [{\citenamefont {Murao}\ \emph {et~al.}(1999)\citenamefont {Murao},
  \citenamefont {Jonathan}, \citenamefont {Plenio},\ and\ \citenamefont
  {Vedral}}]{Murao99}%
  \BibitemOpen
  \bibfield  {author} {\bibinfo {author} {\bibfnamefont {M.}~\bibnamefont
  {Murao}}, \bibinfo {author} {\bibfnamefont {D.}~\bibnamefont {Jonathan}},
  \bibinfo {author} {\bibfnamefont {M.~B.}\ \bibnamefont {Plenio}}, \ and\
  \bibinfo {author} {\bibfnamefont {V.}~\bibnamefont {Vedral}},\ }\href
  {\doibase 10.1103/PhysRevA.59.156} {\bibfield  {journal} {\bibinfo  {journal}
  {Phys. Rev. A}\ }\textbf {\bibinfo {volume} {59}},\ \bibinfo {pages} {156}
  (\bibinfo {year} {1999})}\BibitemShut {NoStop}%
\bibitem [{\citenamefont {Markham}\ and\ \citenamefont
  {Sanders}(2008)}]{Markham08}%
  \BibitemOpen
  \bibfield  {author} {\bibinfo {author} {\bibfnamefont {D.}~\bibnamefont
  {Markham}}\ and\ \bibinfo {author} {\bibfnamefont {B.~C.}\ \bibnamefont
  {Sanders}},\ }\href {\doibase 10.1103/PhysRevA.78.042309} {\bibfield
  {journal} {\bibinfo  {journal} {Phys. Rev. A}\ }\textbf {\bibinfo {volume}
  {78}},\ \bibinfo {pages} {042309} (\bibinfo {year} {2008})}\BibitemShut
  {NoStop}%
\bibitem [{\citenamefont {Bennett}\ and\ \citenamefont
  {BRassard}(1984)}]{BB84}%
  \BibitemOpen
  \bibfield  {author} {\bibinfo {author} {\bibfnamefont {C.}~\bibnamefont
  {Bennett}}\ and\ \bibinfo {author} {\bibfnamefont {G.}~\bibnamefont
  {BRassard}},\ }\href@noop {} {\bibfield  {journal} {\bibinfo  {journal}
  {Proceedings of IEEE International Conference on Computers, Systems and
  Signal Processing}\ ,\ \bibinfo {pages} {175}} (\bibinfo {year}
  {1984})}\BibitemShut {NoStop}%
\bibitem [{\citenamefont {{Chen}}\ and\ \citenamefont {{Lo}}(2004)}]{Chen04}%
  \BibitemOpen
  \bibfield  {author} {\bibinfo {author} {\bibfnamefont {K.}~\bibnamefont
  {{Chen}}}\ and\ \bibinfo {author} {\bibfnamefont {H.-K.}\ \bibnamefont
  {{Lo}}},\ }\href@noop {} {\bibfield  {journal} {\bibinfo  {journal} {eprint
  arXiv:quant-ph/0404133}\ } (\bibinfo {year} {2004})},\ \Eprint
  {http://arxiv.org/abs/quant-ph/0404133} {quant-ph/0404133} \BibitemShut
  {NoStop}%
\bibitem [{\citenamefont {D\"ur}\ \emph {et~al.}(2005)\citenamefont {D\"ur},
  \citenamefont {Calsamiglia},\ and\ \citenamefont {Briegel}}]{Duer05}%
  \BibitemOpen
  \bibfield  {author} {\bibinfo {author} {\bibfnamefont {W.}~\bibnamefont
  {D\"ur}}, \bibinfo {author} {\bibfnamefont {J.}~\bibnamefont {Calsamiglia}},
  \ and\ \bibinfo {author} {\bibfnamefont {H.-J.}\ \bibnamefont {Briegel}},\
  }\href {\doibase 10.1103/PhysRevA.71.042336} {\bibfield  {journal} {\bibinfo
  {journal} {Phys. Rev. A}\ }\textbf {\bibinfo {volume} {71}},\ \bibinfo
  {pages} {042336} (\bibinfo {year} {2005})}\BibitemShut {NoStop}%
\bibitem [{\citenamefont {Mermin}(1990)}]{Mermin1990}%
  \BibitemOpen
  \bibfield  {author} {\bibinfo {author} {\bibfnamefont {N.}~\bibnamefont
  {Mermin}},\ }\href@noop {} {\bibfield  {journal} {\bibinfo  {journal} {Phys.
  Rev. Lett.}\ }\textbf {\bibinfo {volume} {65}},\ \bibinfo {pages} {1838}
  (\bibinfo {year} {1990})}\BibitemShut {NoStop}%
\bibitem [{\citenamefont {Briegel}\ \emph {et~al.}(1998)\citenamefont
  {Briegel}, \citenamefont {D\"ur}, \citenamefont {Cirac},\ and\ \citenamefont
  {Zoller}}]{Briegel98}%
  \BibitemOpen
  \bibfield  {author} {\bibinfo {author} {\bibfnamefont {H.-J.}\ \bibnamefont
  {Briegel}}, \bibinfo {author} {\bibfnamefont {W.}~\bibnamefont {D\"ur}},
  \bibinfo {author} {\bibfnamefont {J.~I.}\ \bibnamefont {Cirac}}, \ and\
  \bibinfo {author} {\bibfnamefont {P.}~\bibnamefont {Zoller}},\ }\href
  {\doibase 10.1103/PhysRevLett.81.5932} {\bibfield  {journal} {\bibinfo
  {journal} {Phys. Rev. Lett.}\ }\textbf {\bibinfo {volume} {81}},\ \bibinfo
  {pages} {5932} (\bibinfo {year} {1998})}\BibitemShut {NoStop}%
\bibitem [{\citenamefont {D\"ur}\ \emph {et~al.}(1999)\citenamefont {D\"ur},
  \citenamefont {Briegel}, \citenamefont {Cirac},\ and\ \citenamefont
  {Zoller}}]{Duer99}%
  \BibitemOpen
  \bibfield  {author} {\bibinfo {author} {\bibfnamefont {W.}~\bibnamefont
  {D\"ur}}, \bibinfo {author} {\bibfnamefont {H.-J.}\ \bibnamefont {Briegel}},
  \bibinfo {author} {\bibfnamefont {J.~I.}\ \bibnamefont {Cirac}}, \ and\
  \bibinfo {author} {\bibfnamefont {P.}~\bibnamefont {Zoller}},\ }\href
  {\doibase 10.1103/PhysRevA.59.169} {\bibfield  {journal} {\bibinfo  {journal}
  {Phys. Rev. A}\ }\textbf {\bibinfo {volume} {59}},\ \bibinfo {pages} {169}
  (\bibinfo {year} {1999})}\BibitemShut {NoStop}%
\bibitem [{\citenamefont {Duan}\ \emph {et~al.}()\citenamefont {Duan},
  \citenamefont {Lukin}, \citenamefont {Cirac},\ and\ \citenamefont
  {Zoller}}]{Duan01}%
  \BibitemOpen
  \bibfield  {author} {\bibinfo {author} {\bibfnamefont {L.-M.}\ \bibnamefont
  {Duan}}, \bibinfo {author} {\bibfnamefont {M.~D.}\ \bibnamefont {Lukin}},
  \bibinfo {author} {\bibfnamefont {J.~I.}\ \bibnamefont {Cirac}}, \ and\
  \bibinfo {author} {\bibfnamefont {P.}~\bibnamefont {Zoller}},\ }\href
  {\doibase 10.1038/35106500} {\bibfield  {journal} {\bibinfo  {journal}
  {Nature}\ }\textbf {\bibinfo {volume} {414}},\ \bibinfo {pages}
  {413}}\BibitemShut {NoStop}%
\bibitem [{\citenamefont {van Loock}\ \emph {et~al.}(2006)\citenamefont {van
  Loock}, \citenamefont {Ladd}, \citenamefont {Sanaka}, \citenamefont
  {Yamaguchi}, \citenamefont {Nemoto}, \citenamefont {Munro},\ and\
  \citenamefont {Yamamoto}}]{vanLoock06}%
  \BibitemOpen
  \bibfield  {author} {\bibinfo {author} {\bibfnamefont {P.}~\bibnamefont {van
  Loock}}, \bibinfo {author} {\bibfnamefont {T.~D.}\ \bibnamefont {Ladd}},
  \bibinfo {author} {\bibfnamefont {K.}~\bibnamefont {Sanaka}}, \bibinfo
  {author} {\bibfnamefont {F.}~\bibnamefont {Yamaguchi}}, \bibinfo {author}
  {\bibfnamefont {K.}~\bibnamefont {Nemoto}}, \bibinfo {author} {\bibfnamefont
  {W.~J.}\ \bibnamefont {Munro}}, \ and\ \bibinfo {author} {\bibfnamefont
  {Y.}~\bibnamefont {Yamamoto}},\ }\href {\doibase
  10.1103/PhysRevLett.96.240501} {\bibfield  {journal} {\bibinfo  {journal}
  {Phys. Rev. Lett.}\ }\textbf {\bibinfo {volume} {96}},\ \bibinfo {pages}
  {240501} (\bibinfo {year} {2006})}\BibitemShut {NoStop}%
\bibitem [{\citenamefont {Zwerger}\ \emph {et~al.}(2012)\citenamefont
  {Zwerger}, \citenamefont {D\"ur},\ and\ \citenamefont {Briegel}}]{Zwerger12}%
  \BibitemOpen
  \bibfield  {author} {\bibinfo {author} {\bibfnamefont {M.}~\bibnamefont
  {Zwerger}}, \bibinfo {author} {\bibfnamefont {W.}~\bibnamefont {D\"ur}}, \
  and\ \bibinfo {author} {\bibfnamefont {H.~J.}\ \bibnamefont {Briegel}},\
  }\href {\doibase 10.1103/PhysRevA.85.062326} {\bibfield  {journal} {\bibinfo
  {journal} {Phys. Rev. A}\ }\textbf {\bibinfo {volume} {85}},\ \bibinfo
  {pages} {062326} (\bibinfo {year} {2012})}\BibitemShut {NoStop}%
\bibitem [{\citenamefont {{Knill}}\ and\ \citenamefont
  {{Laflamme}}(1996)}]{Knill96}%
  \BibitemOpen
  \bibfield  {author} {\bibinfo {author} {\bibfnamefont {E.}~\bibnamefont
  {{Knill}}}\ and\ \bibinfo {author} {\bibfnamefont {R.}~\bibnamefont
  {{Laflamme}}},\ }\href@noop {} {\bibfield  {journal} {\bibinfo  {journal}
  {ArXiv e-prints}\ } (\bibinfo {year} {1996})},\ \Eprint
  {http://arxiv.org/abs/quant-ph/9608012} {quant-ph/9608012} \BibitemShut
  {NoStop}%
\bibitem [{\citenamefont {Jiang}\ \emph {et~al.}(2009)\citenamefont {Jiang},
  \citenamefont {Taylor}, \citenamefont {Nemoto}, \citenamefont {Munro},
  \citenamefont {Van~Meter},\ and\ \citenamefont {Lukin}}]{Jiang09}%
  \BibitemOpen
  \bibfield  {author} {\bibinfo {author} {\bibfnamefont {L.}~\bibnamefont
  {Jiang}}, \bibinfo {author} {\bibfnamefont {J.~M.}\ \bibnamefont {Taylor}},
  \bibinfo {author} {\bibfnamefont {K.}~\bibnamefont {Nemoto}}, \bibinfo
  {author} {\bibfnamefont {W.~J.}\ \bibnamefont {Munro}}, \bibinfo {author}
  {\bibfnamefont {R.}~\bibnamefont {Van~Meter}}, \ and\ \bibinfo {author}
  {\bibfnamefont {M.~D.}\ \bibnamefont {Lukin}},\ }\href {\doibase
  10.1103/PhysRevA.79.032325} {\bibfield  {journal} {\bibinfo  {journal} {Phys.
  Rev. A}\ }\textbf {\bibinfo {volume} {79}},\ \bibinfo {pages} {032325}
  (\bibinfo {year} {2009})}\BibitemShut {NoStop}%
\bibitem [{\citenamefont {Fowler}\ \emph {et~al.}(2010)\citenamefont {Fowler},
  \citenamefont {Wang}, \citenamefont {Hill}, \citenamefont {Ladd},
  \citenamefont {Van~Meter},\ and\ \citenamefont {Hollenberg}}]{Fowler10}%
  \BibitemOpen
  \bibfield  {author} {\bibinfo {author} {\bibfnamefont {A.~G.}\ \bibnamefont
  {Fowler}}, \bibinfo {author} {\bibfnamefont {D.~S.}\ \bibnamefont {Wang}},
  \bibinfo {author} {\bibfnamefont {C.~D.}\ \bibnamefont {Hill}}, \bibinfo
  {author} {\bibfnamefont {T.~D.}\ \bibnamefont {Ladd}}, \bibinfo {author}
  {\bibfnamefont {R.}~\bibnamefont {Van~Meter}}, \ and\ \bibinfo {author}
  {\bibfnamefont {L.~C.~L.}\ \bibnamefont {Hollenberg}},\ }\href {\doibase
  10.1103/PhysRevLett.104.180503} {\bibfield  {journal} {\bibinfo  {journal}
  {Phys. Rev. Lett.}\ }\textbf {\bibinfo {volume} {104}},\ \bibinfo {pages}
  {180503} (\bibinfo {year} {2010})}\BibitemShut {NoStop}%
\bibitem [{\citenamefont {Muralidharan}\ \emph {et~al.}(2014)\citenamefont
  {Muralidharan}, \citenamefont {Kim}, \citenamefont {L\"utkenhaus},
  \citenamefont {Lukin},\ and\ \citenamefont {Jiang}}]{Muralidharan14}%
  \BibitemOpen
  \bibfield  {author} {\bibinfo {author} {\bibfnamefont {S.}~\bibnamefont
  {Muralidharan}}, \bibinfo {author} {\bibfnamefont {J.}~\bibnamefont {Kim}},
  \bibinfo {author} {\bibfnamefont {N.}~\bibnamefont {L\"utkenhaus}}, \bibinfo
  {author} {\bibfnamefont {M.~D.}\ \bibnamefont {Lukin}}, \ and\ \bibinfo
  {author} {\bibfnamefont {L.}~\bibnamefont {Jiang}},\ }\href {\doibase
  10.1103/PhysRevLett.112.250501} {\bibfield  {journal} {\bibinfo  {journal}
  {Phys. Rev. Lett.}\ }\textbf {\bibinfo {volume} {112}},\ \bibinfo {pages}
  {250501} (\bibinfo {year} {2014})}\BibitemShut {NoStop}%
\bibitem [{\citenamefont {Cory}\ \emph {et~al.}(1998)\citenamefont {Cory},
  \citenamefont {Price}, \citenamefont {Maas}, \citenamefont {Knill},
  \citenamefont {Laflamme}, \citenamefont {Zurek}, \citenamefont {Havel},\ and\
  \citenamefont {Somaroo}}]{Cory98}%
  \BibitemOpen
  \bibfield  {author} {\bibinfo {author} {\bibfnamefont {D.~G.}\ \bibnamefont
  {Cory}}, \bibinfo {author} {\bibfnamefont {M.~D.}\ \bibnamefont {Price}},
  \bibinfo {author} {\bibfnamefont {W.}~\bibnamefont {Maas}}, \bibinfo {author}
  {\bibfnamefont {E.}~\bibnamefont {Knill}}, \bibinfo {author} {\bibfnamefont
  {R.}~\bibnamefont {Laflamme}}, \bibinfo {author} {\bibfnamefont {W.~H.}\
  \bibnamefont {Zurek}}, \bibinfo {author} {\bibfnamefont {T.~F.}\ \bibnamefont
  {Havel}}, \ and\ \bibinfo {author} {\bibfnamefont {S.~S.}\ \bibnamefont
  {Somaroo}},\ }\href {\doibase 10.1103/PhysRevLett.81.2152} {\bibfield
  {journal} {\bibinfo  {journal} {Phys. Rev. Lett.}\ }\textbf {\bibinfo
  {volume} {81}},\ \bibinfo {pages} {2152} (\bibinfo {year}
  {1998})}\BibitemShut {NoStop}%
\bibitem [{\citenamefont {Yuan}\ \emph {et~al.}(2008)\citenamefont {Yuan},
  \citenamefont {Chen}, \citenamefont {Zhao}, \citenamefont {Chen},
  \citenamefont {Schmiedmayer},\ and\ \citenamefont {Pan}}]{Yuan08}%
  \BibitemOpen
  \bibfield  {author} {\bibinfo {author} {\bibfnamefont {Z.-S.}\ \bibnamefont
  {Yuan}}, \bibinfo {author} {\bibfnamefont {Y.-A.}\ \bibnamefont {Chen}},
  \bibinfo {author} {\bibfnamefont {B.}~\bibnamefont {Zhao}}, \bibinfo {author}
  {\bibfnamefont {S.}~\bibnamefont {Chen}}, \bibinfo {author} {\bibfnamefont
  {J.}~\bibnamefont {Schmiedmayer}}, \ and\ \bibinfo {author} {\bibfnamefont
  {J.-W.}\ \bibnamefont {Pan}},\ }\href {\doibase 10.1038/nature07241}
  {\bibfield  {journal} {\bibinfo  {journal} {Nature}\ }\textbf {\bibinfo
  {volume} {454}},\ \bibinfo {pages} {1098} (\bibinfo {year}
  {2008})}\BibitemShut {NoStop}%
\bibitem [{\citenamefont {Kimble}(2008)}]{Kimble08}%
  \BibitemOpen
  \bibfield  {author} {\bibinfo {author} {\bibfnamefont {H.~J.}\ \bibnamefont
  {Kimble}},\ }\href {\doibase 10.1038/nature07127} {\bibfield  {journal}
  {\bibinfo  {journal} {Nature}\ }\textbf {\bibinfo {volume} {453}},\ \bibinfo
  {pages} {1023} (\bibinfo {year} {2008})}\BibitemShut {NoStop}%
\bibitem [{\citenamefont {Sangouard}\ \emph {et~al.}(2011)\citenamefont
  {Sangouard}, \citenamefont {Simon}, \citenamefont {de~Riedmatten},\ and\
  \citenamefont {Gisin}}]{Sangouard11}%
  \BibitemOpen
  \bibfield  {author} {\bibinfo {author} {\bibfnamefont {N.}~\bibnamefont
  {Sangouard}}, \bibinfo {author} {\bibfnamefont {C.}~\bibnamefont {Simon}},
  \bibinfo {author} {\bibfnamefont {H.}~\bibnamefont {de~Riedmatten}}, \ and\
  \bibinfo {author} {\bibfnamefont {N.}~\bibnamefont {Gisin}},\ }\href
  {\doibase 10.1103/RevModPhys.83.33} {\bibfield  {journal} {\bibinfo
  {journal} {Rev. Mod. Phys.}\ }\textbf {\bibinfo {volume} {83}},\ \bibinfo
  {pages} {33} (\bibinfo {year} {2011})}\BibitemShut {NoStop}%
\bibitem [{\citenamefont {Schindler}\ \emph {et~al.}(2011)\citenamefont
  {Schindler}, \citenamefont {Barreiro}, \citenamefont {Monz}, \citenamefont
  {Nebendahl}, \citenamefont {Nigg}, \citenamefont {Chwalla}, \citenamefont
  {Hennrich},\ and\ \citenamefont {Blatt}}]{Schindler11}%
  \BibitemOpen
  \bibfield  {author} {\bibinfo {author} {\bibfnamefont {P.}~\bibnamefont
  {Schindler}}, \bibinfo {author} {\bibfnamefont {J.~T.}\ \bibnamefont
  {Barreiro}}, \bibinfo {author} {\bibfnamefont {T.}~\bibnamefont {Monz}},
  \bibinfo {author} {\bibfnamefont {V.}~\bibnamefont {Nebendahl}}, \bibinfo
  {author} {\bibfnamefont {D.}~\bibnamefont {Nigg}}, \bibinfo {author}
  {\bibfnamefont {M.}~\bibnamefont {Chwalla}}, \bibinfo {author} {\bibfnamefont
  {M.}~\bibnamefont {Hennrich}}, \ and\ \bibinfo {author} {\bibfnamefont
  {R.}~\bibnamefont {Blatt}},\ }\href {\doibase 10.1126/science.1203329}
  {\bibfield  {journal} {\bibinfo  {journal} {Science}\ }\textbf {\bibinfo
  {volume} {332}},\ \bibinfo {pages} {1059} (\bibinfo {year}
  {2011})}\BibitemShut {NoStop}%
\bibitem [{\citenamefont {Usmani}\ \emph {et~al.}(2012)\citenamefont {Usmani},
  \citenamefont {Clausen}, \citenamefont {Bussieres}, \citenamefont
  {Sangouard}, \citenamefont {Afzelius},\ and\ \citenamefont
  {Gisin}}]{Clausen12}%
  \BibitemOpen
  \bibfield  {author} {\bibinfo {author} {\bibfnamefont {I.}~\bibnamefont
  {Usmani}}, \bibinfo {author} {\bibfnamefont {C.}~\bibnamefont {Clausen}},
  \bibinfo {author} {\bibfnamefont {F.}~\bibnamefont {Bussieres}}, \bibinfo
  {author} {\bibfnamefont {N.}~\bibnamefont {Sangouard}}, \bibinfo {author}
  {\bibfnamefont {M.}~\bibnamefont {Afzelius}}, \ and\ \bibinfo {author}
  {\bibfnamefont {N.}~\bibnamefont {Gisin}},\ }\href {\doibase
  10.1038/nphoton.2012.34} {\bibfield  {journal} {\bibinfo  {journal} {Nat.
  Photon.}\ }\textbf {\bibinfo {volume} {6}},\ \bibinfo {pages} {234} (\bibinfo
  {year} {2012})}\BibitemShut {NoStop}%
\bibitem [{\citenamefont {{Hensen}}\ \emph {et~al.}(2015)\citenamefont
  {{Hensen}}, \citenamefont {{Bernien}}, \citenamefont {{Dr{\'e}au}},
  \citenamefont {{Reiserer}}, \citenamefont {{Kalb}}, \citenamefont {{Blok}},
  \citenamefont {{Ruitenberg}}, \citenamefont {{Vermeulen}}, \citenamefont
  {{Schouten}}, \citenamefont {{Abell{\'a}n}}, \citenamefont {{Amaya}},
  \citenamefont {{Pruneri}}, \citenamefont {{Mitchell}}, \citenamefont
  {{Markham}}, \citenamefont {{Twitchen}}, \citenamefont {{Elkouss}},
  \citenamefont {{Wehner}}, \citenamefont {{Taminiau}},\ and\ \citenamefont
  {{Hanson}}}]{Hensen15}%
  \BibitemOpen
  \bibfield  {author} {\bibinfo {author} {\bibfnamefont {B.}~\bibnamefont
  {{Hensen}}}, \bibinfo {author} {\bibfnamefont {H.}~\bibnamefont {{Bernien}}},
  \bibinfo {author} {\bibfnamefont {A.~E.}\ \bibnamefont {{Dr{\'e}au}}},
  \bibinfo {author} {\bibfnamefont {A.}~\bibnamefont {{Reiserer}}}, \bibinfo
  {author} {\bibfnamefont {N.}~\bibnamefont {{Kalb}}}, \bibinfo {author}
  {\bibfnamefont {M.~S.}\ \bibnamefont {{Blok}}}, \bibinfo {author}
  {\bibfnamefont {J.}~\bibnamefont {{Ruitenberg}}}, \bibinfo {author}
  {\bibfnamefont {R.~F.~L.}\ \bibnamefont {{Vermeulen}}}, \bibinfo {author}
  {\bibfnamefont {R.~N.}\ \bibnamefont {{Schouten}}}, \bibinfo {author}
  {\bibfnamefont {C.}~\bibnamefont {{Abell{\'a}n}}}, \bibinfo {author}
  {\bibfnamefont {W.}~\bibnamefont {{Amaya}}}, \bibinfo {author} {\bibfnamefont
  {V.}~\bibnamefont {{Pruneri}}}, \bibinfo {author} {\bibfnamefont {M.~W.}\
  \bibnamefont {{Mitchell}}}, \bibinfo {author} {\bibfnamefont
  {M.}~\bibnamefont {{Markham}}}, \bibinfo {author} {\bibfnamefont {D.~J.}\
  \bibnamefont {{Twitchen}}}, \bibinfo {author} {\bibfnamefont
  {D.}~\bibnamefont {{Elkouss}}}, \bibinfo {author} {\bibfnamefont
  {S.}~\bibnamefont {{Wehner}}}, \bibinfo {author} {\bibfnamefont {T.~H.}\
  \bibnamefont {{Taminiau}}}, \ and\ \bibinfo {author} {\bibfnamefont
  {R.}~\bibnamefont {{Hanson}}},\ }\href@noop {} {\bibfield  {journal}
  {\bibinfo  {journal} {ArXiv e-prints:quant-ph/1508.05949}\ } (\bibinfo {year}
  {2015})},\ \Eprint {http://arxiv.org/abs/1508.05949} {arXiv:1508.05949
  [quant-ph]} \BibitemShut {NoStop}%
\bibitem [{\citenamefont {Peev}\ \emph {et~al.}(2009)\citenamefont {Peev},
  \citenamefont {Pacher}, \citenamefont {Alléaume}, \citenamefont {Barreiro},
  \citenamefont {J}, \citenamefont {Boxleitner}, \citenamefont {Debuisschert},
  \citenamefont {Diamanti}, \citenamefont {Dianati}, \citenamefont {Dynes},
  \citenamefont {Fasel}, \citenamefont {Fossier}, \citenamefont {M},
  \citenamefont {Gautier}, \citenamefont {Gay}, \citenamefont {Gisin},
  \citenamefont {Grangier}, \citenamefont {Happe}, \citenamefont {Hasani},
  \citenamefont {Hentschel}, \citenamefont {Hübel}, \citenamefont {Humer},
  \citenamefont {Länger}, \citenamefont {Legré}, \citenamefont {Lieger},
  \citenamefont {Lodewyck}, \citenamefont {Lorünser}, \citenamefont
  {Lütkenhaus}, \citenamefont {Marhold}, \citenamefont {Matyus}, \citenamefont
  {Maurhart}, \citenamefont {Monat}, \citenamefont {Nauerth}, \citenamefont
  {Page}, \citenamefont {Poppe}, \citenamefont {Querasser}, \citenamefont
  {Ribordy}, \citenamefont {Robyr}, \citenamefont {Salvail}, \citenamefont
  {Sharpe}, \citenamefont {Shields}, \citenamefont {Stucki}, \citenamefont
  {Suda}, \citenamefont {Tamas}, \citenamefont {Themel}, \citenamefont {Thew},
  \citenamefont {Thoma}, \citenamefont {Treiber}, \citenamefont {Trinkler},
  \citenamefont {Tualle-Brouri}, \citenamefont {Vannel}, \citenamefont
  {Walenta}, \citenamefont {Weier}, \citenamefont {Weinfurter}, \citenamefont
  {Wimberger}, \citenamefont {Yuan}, \citenamefont {Zbinden},\ and\
  \citenamefont {Zeilinger}}]{SECOQC09}%
  \BibitemOpen
  \bibfield  {author} {\bibinfo {author} {\bibfnamefont {M.}~\bibnamefont
  {Peev}}, \bibinfo {author} {\bibfnamefont {C.}~\bibnamefont {Pacher}},
  \bibinfo {author} {\bibfnamefont {R.}~\bibnamefont {Alléaume}}, \bibinfo
  {author} {\bibfnamefont {C.}~\bibnamefont {Barreiro}}, \bibinfo {author}
  {\bibfnamefont {B.}~\bibnamefont {J}}, \bibinfo {author} {\bibfnamefont
  {W.}~\bibnamefont {Boxleitner}}, \bibinfo {author} {\bibfnamefont
  {T.}~\bibnamefont {Debuisschert}}, \bibinfo {author} {\bibfnamefont
  {E.}~\bibnamefont {Diamanti}}, \bibinfo {author} {\bibfnamefont
  {M.}~\bibnamefont {Dianati}}, \bibinfo {author} {\bibfnamefont {J.~F.}\
  \bibnamefont {Dynes}}, \bibinfo {author} {\bibfnamefont {S.}~\bibnamefont
  {Fasel}}, \bibinfo {author} {\bibfnamefont {S.}~\bibnamefont {Fossier}},
  \bibinfo {author} {\bibfnamefont {F.}~\bibnamefont {M}}, \bibinfo {author}
  {\bibfnamefont {J.-D.}\ \bibnamefont {Gautier}}, \bibinfo {author}
  {\bibfnamefont {O.}~\bibnamefont {Gay}}, \bibinfo {author} {\bibfnamefont
  {N.}~\bibnamefont {Gisin}}, \bibinfo {author} {\bibfnamefont
  {P.}~\bibnamefont {Grangier}}, \bibinfo {author} {\bibfnamefont
  {A.}~\bibnamefont {Happe}}, \bibinfo {author} {\bibfnamefont
  {Y.}~\bibnamefont {Hasani}}, \bibinfo {author} {\bibfnamefont
  {M.}~\bibnamefont {Hentschel}}, \bibinfo {author} {\bibfnamefont
  {H.}~\bibnamefont {Hübel}}, \bibinfo {author} {\bibfnamefont
  {G.}~\bibnamefont {Humer}}, \bibinfo {author} {\bibfnamefont
  {T.}~\bibnamefont {Länger}}, \bibinfo {author} {\bibfnamefont
  {M.}~\bibnamefont {Legré}}, \bibinfo {author} {\bibfnamefont
  {R.}~\bibnamefont {Lieger}}, \bibinfo {author} {\bibfnamefont
  {J.}~\bibnamefont {Lodewyck}}, \bibinfo {author} {\bibfnamefont
  {T.}~\bibnamefont {Lorünser}}, \bibinfo {author} {\bibfnamefont
  {N.}~\bibnamefont {Lütkenhaus}}, \bibinfo {author} {\bibfnamefont
  {A.}~\bibnamefont {Marhold}}, \bibinfo {author} {\bibfnamefont
  {T.}~\bibnamefont {Matyus}}, \bibinfo {author} {\bibfnamefont
  {O.}~\bibnamefont {Maurhart}}, \bibinfo {author} {\bibfnamefont
  {L.}~\bibnamefont {Monat}}, \bibinfo {author} {\bibfnamefont
  {S.}~\bibnamefont {Nauerth}}, \bibinfo {author} {\bibfnamefont {J.-B.}\
  \bibnamefont {Page}}, \bibinfo {author} {\bibfnamefont {A.}~\bibnamefont
  {Poppe}}, \bibinfo {author} {\bibfnamefont {E.}~\bibnamefont {Querasser}},
  \bibinfo {author} {\bibfnamefont {G.}~\bibnamefont {Ribordy}}, \bibinfo
  {author} {\bibfnamefont {S.}~\bibnamefont {Robyr}}, \bibinfo {author}
  {\bibfnamefont {L.}~\bibnamefont {Salvail}}, \bibinfo {author} {\bibfnamefont
  {A.~W.}\ \bibnamefont {Sharpe}}, \bibinfo {author} {\bibfnamefont {A.~J.}\
  \bibnamefont {Shields}}, \bibinfo {author} {\bibfnamefont {D.}~\bibnamefont
  {Stucki}}, \bibinfo {author} {\bibfnamefont {M.}~\bibnamefont {Suda}},
  \bibinfo {author} {\bibfnamefont {C.}~\bibnamefont {Tamas}}, \bibinfo
  {author} {\bibfnamefont {T.}~\bibnamefont {Themel}}, \bibinfo {author}
  {\bibfnamefont {R.~T.}\ \bibnamefont {Thew}}, \bibinfo {author}
  {\bibfnamefont {Y.}~\bibnamefont {Thoma}}, \bibinfo {author} {\bibfnamefont
  {A.}~\bibnamefont {Treiber}}, \bibinfo {author} {\bibfnamefont
  {P.}~\bibnamefont {Trinkler}}, \bibinfo {author} {\bibfnamefont
  {R.}~\bibnamefont {Tualle-Brouri}}, \bibinfo {author} {\bibfnamefont
  {F.}~\bibnamefont {Vannel}}, \bibinfo {author} {\bibfnamefont
  {N.}~\bibnamefont {Walenta}}, \bibinfo {author} {\bibfnamefont
  {H.}~\bibnamefont {Weier}}, \bibinfo {author} {\bibfnamefont
  {H.}~\bibnamefont {Weinfurter}}, \bibinfo {author} {\bibfnamefont
  {I.}~\bibnamefont {Wimberger}}, \bibinfo {author} {\bibfnamefont {Z.~L.}\
  \bibnamefont {Yuan}}, \bibinfo {author} {\bibfnamefont {H.}~\bibnamefont
  {Zbinden}}, \ and\ \bibinfo {author} {\bibfnamefont {A.}~\bibnamefont
  {Zeilinger}},\ }\href {http://stacks.iop.org/1367-2630/11/i=7/a=075001}
  {\bibfield  {journal} {\bibinfo  {journal} {New Journal of Physics}\ }\textbf
  {\bibinfo {volume} {11}},\ \bibinfo {pages} {075001} (\bibinfo {year}
  {2009})}\BibitemShut {NoStop}%
\bibitem [{\citenamefont {Elliott}(2002)}]{Elliott02}%
  \BibitemOpen
  \bibfield  {author} {\bibinfo {author} {\bibfnamefont {C.}~\bibnamefont
  {Elliott}},\ }\href {http://stacks.iop.org/1367-2630/4/i=1/a=346} {\bibfield
  {journal} {\bibinfo  {journal} {New Journal of Physics}\ }\textbf {\bibinfo
  {volume} {4}},\ \bibinfo {pages} {46} (\bibinfo {year} {2002})}\BibitemShut
  {NoStop}%
\bibitem [{\citenamefont {Acin}\ \emph {et~al.}(2007)\citenamefont {Acin},
  \citenamefont {Cirac},\ and\ \citenamefont {Lewenstein}}]{Acin07}%
  \BibitemOpen
  \bibfield  {author} {\bibinfo {author} {\bibfnamefont {A.}~\bibnamefont
  {Acin}}, \bibinfo {author} {\bibfnamefont {J.~I.}\ \bibnamefont {Cirac}}, \
  and\ \bibinfo {author} {\bibfnamefont {M.}~\bibnamefont {Lewenstein}},\
  }\href {\doibase 10.1038/nphys549} {\bibfield  {journal} {\bibinfo  {journal}
  {Nat. Phys.}\ }\textbf {\bibinfo {volume} {3}},\ \bibinfo {pages} {256}
  (\bibinfo {year} {2007})}\BibitemShut {NoStop}%
\bibitem [{\citenamefont {Van~Meter}\ \emph {et~al.}(2013)\citenamefont
  {Van~Meter}, \citenamefont {Satoh}, \citenamefont {Ladd}, \citenamefont
  {Munro},\ and\ \citenamefont {Nemoto}}]{VanMeter13}%
  \BibitemOpen
  \bibfield  {author} {\bibinfo {author} {\bibfnamefont {R.}~\bibnamefont
  {Van~Meter}}, \bibinfo {author} {\bibfnamefont {T.}~\bibnamefont {Satoh}},
  \bibinfo {author} {\bibfnamefont {T.}~\bibnamefont {Ladd}}, \bibinfo {author}
  {\bibfnamefont {W.}~\bibnamefont {Munro}}, \ and\ \bibinfo {author}
  {\bibfnamefont {K.}~\bibnamefont {Nemoto}},\ }\href {\doibase
  10.1007/s13119-013-0026-2} {\bibfield  {journal} {\bibinfo  {journal}
  {Networking Science}\ }\textbf {\bibinfo {volume} {3}},\ \bibinfo {pages}
  {82} (\bibinfo {year} {2013})}\BibitemShut {NoStop}%
\bibitem [{\citenamefont {Perseguers}\ \emph {et~al.}(2013)\citenamefont
  {Perseguers}, \citenamefont {Jr}, \citenamefont {Cavalcanti}, \citenamefont
  {Lewenstein},\ and\ \citenamefont {Aci­n}}]{Perseguers13}%
  \BibitemOpen
  \bibfield  {author} {\bibinfo {author} {\bibfnamefont {S.}~\bibnamefont
  {Perseguers}}, \bibinfo {author} {\bibfnamefont {G.~J.~L.}\ \bibnamefont
  {Jr}}, \bibinfo {author} {\bibfnamefont {D.}~\bibnamefont {Cavalcanti}},
  \bibinfo {author} {\bibfnamefont {M.}~\bibnamefont {Lewenstein}}, \ and\
  \bibinfo {author} {\bibfnamefont {A.}~\bibnamefont {Aci­n}},\ }\href
  {http://stacks.iop.org/0034-4885/76/i=9/a=096001} {\bibfield  {journal}
  {\bibinfo  {journal} {Reports on Progress in Physics}\ }\textbf {\bibinfo
  {volume} {76}},\ \bibinfo {pages} {096001} (\bibinfo {year}
  {2013})}\BibitemShut {NoStop}%
\bibitem [{\citenamefont {{Leung}}\ \emph {et~al.}(2006)\citenamefont
  {{Leung}}, \citenamefont {{Oppenheim}},\ and\ \citenamefont
  {{Winter}}}]{Leung06}%
  \BibitemOpen
  \bibfield  {author} {\bibinfo {author} {\bibfnamefont {D.}~\bibnamefont
  {{Leung}}}, \bibinfo {author} {\bibfnamefont {J.}~\bibnamefont
  {{Oppenheim}}}, \ and\ \bibinfo {author} {\bibfnamefont {A.}~\bibnamefont
  {{Winter}}},\ }\href@noop {} {\bibfield  {journal} {\bibinfo  {journal} {IEEE
  Trans. Inf. Theory}\ }\textbf {\bibinfo {volume} {56}},\ \bibinfo {pages}
  {3478} (\bibinfo {year} {2006})}\BibitemShut {NoStop}%
\bibitem [{\citenamefont {Hayashi}(2007)}]{Hayashi07}%
  \BibitemOpen
  \bibfield  {author} {\bibinfo {author} {\bibfnamefont {M.}~\bibnamefont
  {Hayashi}},\ }\href {\doibase 10.1103/PhysRevA.76.040301} {\bibfield
  {journal} {\bibinfo  {journal} {Phys. Rev. A}\ }\textbf {\bibinfo {volume}
  {76}},\ \bibinfo {pages} {040301} (\bibinfo {year} {2007})}\BibitemShut
  {NoStop}%
\bibitem [{\citenamefont {{Nagayama}}\ \emph {et~al.}(2015)\citenamefont
  {{Nagayama}}, \citenamefont {{Choi}}, \citenamefont {{Devitt}}, \citenamefont
  {{Suzuki}},\ and\ \citenamefont {{Van Meter}}}]{Nagayama15}%
  \BibitemOpen
  \bibfield  {author} {\bibinfo {author} {\bibfnamefont {S.}~\bibnamefont
  {{Nagayama}}}, \bibinfo {author} {\bibfnamefont {B.-S.}\ \bibnamefont
  {{Choi}}}, \bibinfo {author} {\bibfnamefont {S.}~\bibnamefont {{Devitt}}},
  \bibinfo {author} {\bibfnamefont {S.}~\bibnamefont {{Suzuki}}}, \ and\
  \bibinfo {author} {\bibfnamefont {R.}~\bibnamefont {{Van Meter}}},\
  }\href@noop {} {\bibfield  {journal} {\bibinfo  {journal} {ArXiv
  e-prints:quant-ph/1508.04599}\ } (\bibinfo {year} {2015})},\ \Eprint
  {http://arxiv.org/abs/1508.04599} {arXiv:1508.04599 [quant-ph]} \BibitemShut
  {NoStop}%
\bibitem [{\citenamefont {Schlingemann}\ and\ \citenamefont
  {Werner}(2001)}]{Schlingemann01}%
  \BibitemOpen
  \bibfield  {author} {\bibinfo {author} {\bibfnamefont {D.}~\bibnamefont
  {Schlingemann}}\ and\ \bibinfo {author} {\bibfnamefont {R.~F.}\ \bibnamefont
  {Werner}},\ }\href {\doibase 10.1103/PhysRevA.65.012308} {\bibfield
  {journal} {\bibinfo  {journal} {Phys. Rev. A}\ }\textbf {\bibinfo {volume}
  {65}},\ \bibinfo {pages} {012308} (\bibinfo {year} {2001})}\BibitemShut
  {NoStop}%
\bibitem [{\citenamefont {Hein}\ \emph {et~al.}(2004)\citenamefont {Hein},
  \citenamefont {Eisert},\ and\ \citenamefont {Briegel}}]{Hein04}%
  \BibitemOpen
  \bibfield  {author} {\bibinfo {author} {\bibfnamefont {M.}~\bibnamefont
  {Hein}}, \bibinfo {author} {\bibfnamefont {J.}~\bibnamefont {Eisert}}, \ and\
  \bibinfo {author} {\bibfnamefont {H.~J.}\ \bibnamefont {Briegel}},\ }\href
  {\doibase 10.1103/PhysRevA.69.062311} {\bibfield  {journal} {\bibinfo
  {journal} {Phys. Rev. A}\ }\textbf {\bibinfo {volume} {69}},\ \bibinfo
  {pages} {062311} (\bibinfo {year} {2004})}\BibitemShut {NoStop}%
\bibitem [{\citenamefont {Raussendorf}\ \emph {et~al.}(2003)\citenamefont
  {Raussendorf}, \citenamefont {Browne},\ and\ \citenamefont
  {Briegel}}]{Raussendorf03}%
  \BibitemOpen
  \bibfield  {author} {\bibinfo {author} {\bibfnamefont {R.}~\bibnamefont
  {Raussendorf}}, \bibinfo {author} {\bibfnamefont {D.~E.}\ \bibnamefont
  {Browne}}, \ and\ \bibinfo {author} {\bibfnamefont {H.~J.}\ \bibnamefont
  {Briegel}},\ }\href {\doibase 10.1103/PhysRevA.68.022312} {\bibfield
  {journal} {\bibinfo  {journal} {Phys. Rev. A}\ }\textbf {\bibinfo {volume}
  {68}},\ \bibinfo {pages} {022312} (\bibinfo {year} {2003})}\BibitemShut
  {NoStop}%
\bibitem [{\citenamefont {D\"ur}\ \emph {et~al.}(2003)\citenamefont {D\"ur},
  \citenamefont {Aschauer},\ and\ \citenamefont {Briegel}}]{Duer03}%
  \BibitemOpen
  \bibfield  {author} {\bibinfo {author} {\bibfnamefont {W.}~\bibnamefont
  {D\"ur}}, \bibinfo {author} {\bibfnamefont {H.}~\bibnamefont {Aschauer}}, \
  and\ \bibinfo {author} {\bibfnamefont {H.-J.}\ \bibnamefont {Briegel}},\
  }\href {\doibase 10.1103/PhysRevLett.91.107903} {\bibfield  {journal}
  {\bibinfo  {journal} {Phys. Rev. Lett.}\ }\textbf {\bibinfo {volume} {91}},\
  \bibinfo {pages} {107903} (\bibinfo {year} {2003})}\BibitemShut {NoStop}%
\bibitem [{\citenamefont {G\"uhne}\ \emph {et~al.}(2005)\citenamefont
  {G\"uhne}, \citenamefont {T\'oth}, \citenamefont {Hyllus},\ and\
  \citenamefont {Briegel}}]{Guehne05}%
  \BibitemOpen
  \bibfield  {author} {\bibinfo {author} {\bibfnamefont {O.}~\bibnamefont
  {G\"uhne}}, \bibinfo {author} {\bibfnamefont {G.}~\bibnamefont {T\'oth}},
  \bibinfo {author} {\bibfnamefont {P.}~\bibnamefont {Hyllus}}, \ and\ \bibinfo
  {author} {\bibfnamefont {H.~J.}\ \bibnamefont {Briegel}},\ }\href {\doibase
  10.1103/PhysRevLett.95.120405} {\bibfield  {journal} {\bibinfo  {journal}
  {Phys. Rev. Lett.}\ }\textbf {\bibinfo {volume} {95}},\ \bibinfo {pages}
  {120405} (\bibinfo {year} {2005})}\BibitemShut {NoStop}%
\bibitem [{\citenamefont {Wu}\ \emph {et~al.}(2015)\citenamefont {Wu},
  \citenamefont {Kampermann},\ and\ \citenamefont {Bru\ss{}}}]{Wu15}%
  \BibitemOpen
  \bibfield  {author} {\bibinfo {author} {\bibfnamefont {J.-Y.}\ \bibnamefont
  {Wu}}, \bibinfo {author} {\bibfnamefont {H.}~\bibnamefont {Kampermann}}, \
  and\ \bibinfo {author} {\bibfnamefont {D.}~\bibnamefont {Bru\ss{}}},\ }\href
  {\doibase 10.1103/PhysRevA.92.012322} {\bibfield  {journal} {\bibinfo
  {journal} {Phys. Rev. A}\ }\textbf {\bibinfo {volume} {92}},\ \bibinfo
  {pages} {012322} (\bibinfo {year} {2015})}\BibitemShut {NoStop}%
\bibitem [{\citenamefont {Lidar}\ and\ \citenamefont
  {Brun}(2013)}]{LidarBrunQEC13}%
  \BibitemOpen
  \bibfield  {author} {\bibinfo {author} {\bibfnamefont {D.}~\bibnamefont
  {Lidar}}\ and\ \bibinfo {author} {\bibfnamefont {T.}~\bibnamefont {Brun}},\
  }\href@noop {} {\emph {\bibinfo {title} {Quantum Error Correction}}}\
  (\bibinfo  {publisher} {Cambridge University Press},\ \bibinfo {year}
  {2013})\BibitemShut {NoStop}%
\bibitem [{\citenamefont {{Zwerger}}\ \emph {et~al.}(2014)\citenamefont
  {{Zwerger}}, \citenamefont {{Briegel}},\ and\ \citenamefont
  {{D{\"u}r}}}]{Zwerger14}%
  \BibitemOpen
  \bibfield  {author} {\bibinfo {author} {\bibfnamefont {M.}~\bibnamefont
  {{Zwerger}}}, \bibinfo {author} {\bibfnamefont {H.~J.}\ \bibnamefont
  {{Briegel}}}, \ and\ \bibinfo {author} {\bibfnamefont {W.}~\bibnamefont
  {{D{\"u}r}}},\ }\href {\doibase 10.1038/srep05364} {\bibfield  {journal}
  {\bibinfo  {journal} {Scientific Reports}\ }\textbf {\bibinfo {volume} {4}},\
  \bibinfo {eid} {5364} (\bibinfo {year} {2014})},\ \Eprint
  {http://arxiv.org/abs/1308.4561} {arXiv:1308.4561 [quant-ph]} \BibitemShut
  {NoStop}%
\bibitem [{\citenamefont {{Gottesman}}(1997)}]{Gottesman97}%
  \BibitemOpen
  \bibfield  {author} {\bibinfo {author} {\bibfnamefont {D.}~\bibnamefont
  {{Gottesman}}},\ }\emph {\bibinfo {title} {{Stabilizer codes and quantum
  error correction}}},\ \href@noop {} {Ph.D. thesis},\ \bibinfo  {school}
  {California Institute of Technology} (\bibinfo {year} {1997})\BibitemShut
  {NoStop}%
\bibitem [{\citenamefont {Paetznick}\ and\ \citenamefont
  {Reichardt}(2012)}]{Paetznick11}%
  \BibitemOpen
  \bibfield  {author} {\bibinfo {author} {\bibfnamefont {A.}~\bibnamefont
  {Paetznick}}\ and\ \bibinfo {author} {\bibfnamefont {B.}~\bibnamefont
  {Reichardt}},\ }\href@noop {} {\bibfield  {journal} {\bibinfo  {journal}
  {QIC}\ }\textbf {\bibinfo {volume} {12}},\ \bibinfo {pages} {1034} (\bibinfo
  {year} {2012})}\BibitemShut {NoStop}%
\bibitem [{\citenamefont {Golay}(1949)}]{Golay49}%
  \BibitemOpen
  \bibfield  {author} {\bibinfo {author} {\bibfnamefont {M.~J.~E.}\
  \bibnamefont {Golay}},\ }\href@noop {} {\bibfield  {journal} {\bibinfo
  {journal} {Proc. IRE}\ }\textbf {\bibinfo {volume} {37}},\ \bibinfo {pages}
  {657} (\bibinfo {year} {1949})}\BibitemShut {NoStop}%
\bibitem [{\citenamefont {Goethals}(1971)}]{Goethals71}%
  \BibitemOpen
  \bibfield  {author} {\bibinfo {author} {\bibfnamefont {J.-M.}\ \bibnamefont
  {Goethals}},\ }\href {\doibase
  http://dx.doi.org/10.1016/0097-3165(71)90043-4} {\bibfield  {journal}
  {\bibinfo  {journal} {Journal of Combinatorial Theory, Series A}\ }\textbf
  {\bibinfo {volume} {11}},\ \bibinfo {pages} {178 } (\bibinfo {year}
  {1971})}\BibitemShut {NoStop}%
\bibitem [{\citenamefont {Elia}\ and\ \citenamefont {Taricco}(1995)}]{Elia95}%
  \BibitemOpen
  \bibfield  {author} {\bibinfo {author} {\bibfnamefont {M.}~\bibnamefont
  {Elia}}\ and\ \bibinfo {author} {\bibfnamefont {G.}~\bibnamefont {Taricco}},\
  }\href {\doibase 10.1007/BF02997777} {\bibfield  {journal} {\bibinfo
  {journal} {Annales Des Telecommunications}\ }\textbf {\bibinfo {volume}
  {50}},\ \bibinfo {pages} {721} (\bibinfo {year} {1995})}\BibitemShut
  {NoStop}%
\bibitem [{\citenamefont {Nielsen}\ and\ \citenamefont
  {Chuang}(2000)}]{nielsen00}%
  \BibitemOpen
  \bibfield  {author} {\bibinfo {author} {\bibfnamefont {M.}~\bibnamefont
  {Nielsen}}\ and\ \bibinfo {author} {\bibfnamefont {I.}~\bibnamefont
  {Chuang}},\ }\href {http://books.google.de/books?id=65FqEKQOfP8C} {\emph
  {\bibinfo {title} {Quantum Computation and Quantum Information}}},\ Cambridge
  Series on Information and the Natural Sciences\ (\bibinfo  {publisher}
  {Cambridge University Press},\ \bibinfo {year} {2000})\BibitemShut {NoStop}%
\bibitem [{\citenamefont {Scarani}\ \emph {et~al.}(2009)\citenamefont
  {Scarani}, \citenamefont {Bechmann-Pasquinucci}, \citenamefont {Cerf},
  \citenamefont {Du\ifmmode~\check{s}\else \v{s}\fi{}ek}, \citenamefont
  {L\"utkenhaus},\ and\ \citenamefont {Peev}}]{Scarani09}%
  \BibitemOpen
  \bibfield  {author} {\bibinfo {author} {\bibfnamefont {V.}~\bibnamefont
  {Scarani}}, \bibinfo {author} {\bibfnamefont {H.}~\bibnamefont
  {Bechmann-Pasquinucci}}, \bibinfo {author} {\bibfnamefont {N.~J.}\
  \bibnamefont {Cerf}}, \bibinfo {author} {\bibfnamefont {M.}~\bibnamefont
  {Du\ifmmode~\check{s}\else \v{s}\fi{}ek}}, \bibinfo {author} {\bibfnamefont
  {N.}~\bibnamefont {L\"utkenhaus}}, \ and\ \bibinfo {author} {\bibfnamefont
  {M.}~\bibnamefont {Peev}},\ }\href {\doibase 10.1103/RevModPhys.81.1301}
  {\bibfield  {journal} {\bibinfo  {journal} {Rev. Mod. Phys.}\ }\textbf
  {\bibinfo {volume} {81}},\ \bibinfo {pages} {1301} (\bibinfo {year}
  {2009})}\BibitemShut {NoStop}%
\bibitem [{\citenamefont {Abruzzo}\ \emph {et~al.}(2013)\citenamefont
  {Abruzzo}, \citenamefont {Bratzik}, \citenamefont {Bernardes}, \citenamefont
  {Kampermann}, \citenamefont {van Loock},\ and\ \citenamefont
  {Bru\ss{}}}]{Abruzzo13}%
  \BibitemOpen
  \bibfield  {author} {\bibinfo {author} {\bibfnamefont {S.}~\bibnamefont
  {Abruzzo}}, \bibinfo {author} {\bibfnamefont {S.}~\bibnamefont {Bratzik}},
  \bibinfo {author} {\bibfnamefont {N.~K.}\ \bibnamefont {Bernardes}}, \bibinfo
  {author} {\bibfnamefont {H.}~\bibnamefont {Kampermann}}, \bibinfo {author}
  {\bibfnamefont {P.}~\bibnamefont {van Loock}}, \ and\ \bibinfo {author}
  {\bibfnamefont {D.}~\bibnamefont {Bru\ss{}}},\ }\href {\doibase
  10.1103/PhysRevA.87.052315} {\bibfield  {journal} {\bibinfo  {journal} {Phys.
  Rev. A}\ }\textbf {\bibinfo {volume} {87}},\ \bibinfo {pages} {052315}
  (\bibinfo {year} {2013})}\BibitemShut {NoStop}%
\bibitem [{\citenamefont {Chuang}(2000)}]{Chuang00}%
  \BibitemOpen
  \bibfield  {author} {\bibinfo {author} {\bibfnamefont {I.~L.}\ \bibnamefont
  {Chuang}},\ }\href {\doibase 10.1103/PhysRevLett.85.2006} {\bibfield
  {journal} {\bibinfo  {journal} {Phys. Rev. Lett.}\ }\textbf {\bibinfo
  {volume} {85}},\ \bibinfo {pages} {2006} (\bibinfo {year}
  {2000})}\BibitemShut {NoStop}%
\bibitem [{\citenamefont {{K{\'o}m{\'a}r}}\ \emph {et~al.}(2014)\citenamefont
  {{K{\'o}m{\'a}r}}, \citenamefont {{Kessler}}, \citenamefont {{Bishof}},
  \citenamefont {{Jiang}}, \citenamefont {{S{\o}rensen}}, \citenamefont
  {{Ye}},\ and\ \citenamefont {{Lukin}}}]{Komar14}%
  \BibitemOpen
  \bibfield  {author} {\bibinfo {author} {\bibfnamefont {P.}~\bibnamefont
  {{K{\'o}m{\'a}r}}}, \bibinfo {author} {\bibfnamefont {E.~M.}\ \bibnamefont
  {{Kessler}}}, \bibinfo {author} {\bibfnamefont {M.}~\bibnamefont {{Bishof}}},
  \bibinfo {author} {\bibfnamefont {L.}~\bibnamefont {{Jiang}}}, \bibinfo
  {author} {\bibfnamefont {A.~S.}\ \bibnamefont {{S{\o}rensen}}}, \bibinfo
  {author} {\bibfnamefont {J.}~\bibnamefont {{Ye}}}, \ and\ \bibinfo {author}
  {\bibfnamefont {M.~D.}\ \bibnamefont {{Lukin}}},\ }\href {\doibase
  10.1038/nphys3000} {\bibfield  {journal} {\bibinfo  {journal} {Nature
  Physics}\ }\textbf {\bibinfo {volume} {10}},\ \bibinfo {pages} {582}
  (\bibinfo {year} {2014})},\ \Eprint {http://arxiv.org/abs/1310.6045}
  {arXiv:1310.6045 [quant-ph]} \BibitemShut {NoStop}%
\end{thebibliography}
%

\appendix
\section{Logical graph states}\label{sec:logicalgraphstates}
An $[[n,k,d]]$ error correction code encodes $k$ logical qubits into $n>k$ physical qubits. This redundancy guarantees the correction of up to $\frac{d-1}{2}$ single qubit errors or $d-1$ erasures. Note that erasures, marked in the classical data as '?', are not only caused by losses of qubits at that particular position. In the scheme described in the main text, any noticed error on the qubit $i$ or $i-1$ will lead to an erasure of the measurement outcome at position $i$.\\
Let $S$ denote the stabiliser group of the codespace of a stabiliser code. Furthermore let $\bar{X}^{(1)},...,\bar{X}^{(k)},\bar{Z}^{(1)},...,\bar{Z}^{(k)}$ denote logical $X$- and $Z$-operators. These operators are elements of the normaliser of $S$ but not elements of $S$ and fulfill $X^{(i)}Z^{(i)}=-Z^{(i)}X^{(i)}$, while operators on different logical qubits commute. For simplicity we focus on the $k=1$ case and drop the label of the logical operator. 
In analogy to a graph state, we define a logical graph state associated with a graph $G=(V,E)$ as the unique state $\ket{\bar{G}}$ that is stabilised by operators
\begin{equation}
 \bar{g}_i=\bar{X}_i \prod_{\substack{j\in V\\ (i,j)\in E}} \bar{Z}_j.
\end{equation}
This definition can be generalized to $k>1$ which leads to $k$ copies of the graph state.\\
A logical error $e$ occurs when the recovery operation leads to a wrong codeword $e\ket{\bar{G}}\neq \bar{G}$. This can happen if more errors occurred on one block than the code is able to correct. A logical measurement error occurs if $e$ anticommutes with the observable. One assumes that the probability of (unnoticed) errors $f_u$ and erasures $f_n$ are the same for all physical qubits. Then for any quantum error correction code, the probability of a logical error $\bar{f}_u$ is a function of $f_u$ and $f_n$. It is possible to retry the production on a particular pattern of noticed errors. Thus the function $\bar{f}_u(f_u,f_n)$ depends on the error correction code and the strategy of when to abort. Several examples are given in Appendix~\ref{sec:fqbar}.\\
Given the probability of logical measurement errors at each vertex it is straightforward to calculate the logical error rate at the position of the network nodes. For each main stabiliser $S_i$ centered on party $i$, the applied by-product operator $Z_i^{\sum \bar{x}_j}$ depends on all $\bar{X}$-measurement outcomes $\bar{x}_j$ of the qubits included in the main stabiliser (i.e. the qubits on which it acts non-trivially). These are half of the qubits on the links from and to party $i$, i.e. $\frac{1}{2}\sum_{j\in V} w_{ij}$, where $w_{ij}$ is the number of repeater stations on the link $(i,j)$. Even numbers of logical errors cancel each other and the stabiliser error rate is
\begin{align}
e_i=& \tr\left(\rho\frac{\bar{\1}-\bar{g}_i}{2}\right)\\
   =& \sum_n \tr\left( \rho \bar{Z}_i \prod_{\substack{k\in V\\ k\neq i}} {\bar{Z}_k}^{n^{(k)}} \ket{\bar{G}} \bra{\bar{G}} \bar{Z}_i \prod_{\substack{k'\in V\\ k'\neq i}} {\bar{Z}_{k'}}^{n^{(k')}}\right)\\
   =&\Podd\left(\bar{f}_u(f_u,f_n),\frac{1}{2}\sum_{j\in V} w_{ij}\right), \label{eq:ei}
\end{align}
where $n^{(k)}$ is the $k$-th binary digit of $n$ and
\begin{equation}
 \Podd(f,N)=\frac{1}{2}(1-(1-2f)^N). \label{eq:Podd}
\end{equation}
Suppose that the state produced by the quantum network is given by a density matrix $\rho$. An error on any stabiliser implies the production of a state orthogonal to the target state $\ket{G}$. Note that it suffices to consider only $\bar{Z}$ errors, as the effect of $\bar{X}$-errors can be described with $\bar{Z}$ errors due the stabilisers of the graph state. One can thus immediately gain bounds on the fidelity of $\rho$ with respect to the state $\ket{G}$ from the local error rates $e_v$,
\begin{equation}
 1-\sum_{v\in V} e_v \leq \bra{G}\rho \ket{G} \leq 1-\max\{e_v | v\in V\}.
\end{equation}
For the network given in Fig.~\ref{fig:gC}, these bounds evaluate to $94\%\leq \bra{G}\rho\ket{G} \leq 99\%$, for example.
\section{Error propagation through repeater stations}\label{sec:errprop} 
Calderbank-Shor-Steane (CSS) codes allow the transversal implementation of controlled-NOT gates~\cite{LidarBrunQEC13}. A transversal application of a quantum gate on two blocks of an $[[n,k,d]]$ quantum error correction code is the qubit-wise application of the gate, i.e. $n$ gates act in parallel on the $i$-th qubit of block one and the $i$-th qubit of block two, $i=1,2,...,n$, see Fig.~\ref{fig:logicalrepeater}. By  using two alternating codes in the $\bar{X}$- and $\bar{Z}$-basis on alternating qubits, the transversal application of the controlled-NOT gates acts like a logical controlled-phase gate and we can stick to the language of graph states.
Because only transversal gates are applied in the quantum circuit, it suffices to calculate the physical error rate by considering the first qubit of each block only. We start by calculating the unnoticed error rate $f_u$ for a repeater station, i.e. the probability to get a flipped measurement outcome. Notice that two errors of this kind cancel each other. We collect all independent sources of errors that lead to the error pattern under consideration (e.g. ``flipped outcome at position $i$'') into a vector $\vec{p}$. Then the probability for this error is
\begin{equation}
 f_u = \Poddtilde(\vec{p})=\sum_{\substack{n=0\\|n|_H\text{ odd}}}^{2^N-1} \prod_{k=1}^N p_k^{n^{(k)}}
(1-p_k)^{1-n^{(k)}}, \label{eq:fu}
\end{equation}
where $n^{(k)}$ is the $k$-th binary digit of $n$. In case of the repeater stations the error sources correspond to
\begin{align}
 \vec{p}=&\left(
\Podd\left(\frac{f_{P,u}}{2},2\right),
\frac{f_{P,n}+f_{P,u}}{2},\right.\nonumber\\
    &\left.\Podd\left(\frac{f_{G,u}}{2},3\right),
\Podd\left(\frac{f_{T,u}}{2},2\right),
\frac{f_{M,u}}{2}
\right). \label{eq:p}
\end{align}
Here we included noticed/unnoticed errors for preparation ($f_{P,n/u}$), transmission ($f_{T,n/u}$), gates ($f_{G,n/u}$) and measurement ($f_{M,n/u}$). Remember that $\Podd(f,N)$ was defined in equation (\ref{eq:Podd}). In order to identify the processes that contribute to $f_u$ one checks all possible propagations of errors to a $Z$-error on the qubit $i$ under consideration. It is useful to note that only $X$-errors spread to $Z$-errors on adjacent qubits in a $C_Z$ gate.\\
Please note that we have neglected the difference of the physical error rates for qubits at the boundary of a repeater line and ``typical'' qubits. This is motivated by the fact, that in a large-scale quantum network there are many more repeater stations than parties. Incorporating these boundary effects is however straightforward.\\
We assume that noticed errors do not cancel each other. Again we collect all independent sources of the error under consideration. In contrast to the case of $f_u$, $f_n$ denotes the probability that any of these events occurred. Thus the probability for an error of this type is
\begin{equation}
f_n = 1-(1-f_{P,n})^2(1-f_{G,n})^3(1-f_{T,n})^2(1-f_{M,n})^2. \label{eq:fn}
\end{equation}
\section{Logical error rate of some CSS codes}\label{sec:fqbar}
In order to calculate the secret key rate for a specific encoding, we need the error rates on odd and even logical qubits. The decoder assigns a codeword to any word given by the measurement, i.e. to the true
outcomes altered by the error pattern. This recovered codeword is used to calculate the value of the logical observable. If this recovered
value is different from the ``true outcome'' this word of measurement outcomes contributes to the logical error rate. The decoder may
trigger an abort on any error from the set $\mathcal{F}$. The naive approach to calculating this rate thus is
\begin{equation}
\bar{f}_{u,\mathcal{C}}(f_u,f_n)=\frac{1}{k}\sum_{e\not\in\mathcal{F}} f(e) P_e(e), \label{eq:logicalerrorrate}
\end{equation}
where $P_e(e)$ is the probability of the error $e$ and $f(e)$ is the number of logical errors after decoding. The success probability
depends on $\mathcal{F}$ and reads
\begin{equation}
P_{\mathrm{succ}}=\sum_{e\not\in\mathcal{F}} P_e(e) = 1-\sum_{e\in\mathcal{F}} P_e(e).
\end{equation}
Let us consider a decoder that returns the most likely codeword $c$ in the sense that an error pattern $e$ that changes $c$ to the observed
data has maximal probability $P_e(e)$. If this is not unique the decoder chooses any such $c$ with equal probability. We use it for the
7-qubit Steane code and use a fatal error set $\mathcal{F}$ of the form
\begin{equation}
\mathcal{F}=\{e|e \text{ contains more than $n_{\mathrm{max}}$ losses} \},
\end{equation}
where $n_{\mathrm{max}}\in\mathds{N}$, i.e. the protocol is aborted if more than $n_{\mathrm{max}}$ losses occurred. The error rates of the
7-qubit Steane code listed in Table~\ref{tab:steane} were obtained by implementing equation~(\ref{eq:logicalerrorrate}).
\section{Generalization of the error analysis}\label{sec:generalization}
Given a graph $G=(V,E)$ with vertices $V$ and edges $E$. The state $\ket{G}$ is stabilised by the generators of the stabiliser $g_i$
($i=1,...,|V|$).The repeater network that creates the graph state $\ket{G}$ is obtained by replacing each edge $(i,j)$ in $E$
by a line graph with $w_{ij}$ additional vertices (the repeater stations). Let us assume that $w_{ij}$ is even, for simplicity. 
All repeater stations are measured in the $\bar{X}$ basis. This projects onto a state that is stabilised by the $g_i$ up to byproduct
operators
that depend on the measurement outcomes. Thus after application of the byproduct operators $g_i \ket{G}=\ket{G}$ holds. A flip of one
measurement outcome on the $i$-th main stabiliser ($g_i$ connected by chains of $X$-operators) leads to $g_i \ket{G}=-\ket{G}$. The same
holds for $\bar{X}$ errors on the neighbors of party $i$ or a $Z$ error on the qubit of party $i$. The corresponding error probability is
$f_i$. We denote the probability for the wrong sign in the stabiliser equation of $g_i$ by $e_i$. It is
\begin{equation}
 e_i=\Poddtilde\left(\left( \Podd\left(\bar{f}_u,\sum_j w_{ij}\right),f_i,f_j|(i,j)\in E\right)\right).
\end{equation}
In analogy to equations~(14) and (16) of the article one can estimate the error rate 
\begin{equation}
 \begin{aligned}
f_{i,u}=&\Poddtilde
\left(\left(\Podd\left(\frac{f_{P,u}}{2},1+\deg^-(i)\right),\right.\right.\\
    &\Podd\left(\frac{f_{P,n}+f_{P,u}}{2},\deg^+(i)\right),\\
    &\Podd\left(\frac{f_{G,u}}{2},1+\deg(i)\right),\\
    &\Podd\left(\left.\left.\frac{f_{T,u}}{2},1+\deg^-(i)\right),\frac{f_{M,u}}{2}\right)\right)
\end{aligned} \label{eq:fiu}\\
\end{equation}
and
\begin{equation}
\begin{aligned}
 f_{i,n} =&1-(1-f_{P,n})^{1+\deg^-(i)}(1-f_{G,n})^{1+\deg(i)}\\
 &(1-f_{T,n})^{1+\deg^-(i)}(1-f_{M,n})^{1+\deg^-(i)}, 
\end{aligned}\label{eq:fin}
\end{equation}
where $\deg(i)$, $\deg^-(i)$, and $\deg^+(i)$ are the degree, in-degree, and out-degree of vertex $i$, respectively. From these physical
error rates one can calculate the logical error rate in analogy to $\bar{f}_u$, i.e. $\bar{f}_i=\bar{f}_u(f_u=f_{i,u},f_n=f_{i,n})$.
\begin{table*}[htp]
\caption{Logical error rates of the Steane code}\label{tab:steane}
\begin{tabular}{rl}
& 7-qubit Steane code ($n_{\mathrm{max}}=0$)\\
$\bar{f}_u(f_u,f_n)=$ &$\left(f_n-1\right)^7 f_u^2 \left(48 f_u^5-168 f_u^4+252 f_u^3-210 f_u^2+98 f_u-21\right)$ \\
$P_{\mathrm{succ}}(f_n)=$ & $\left(1-f_n\right)^7$\\[1ex]
&7-qubit Steane code ($n_{\mathrm{max}}=1$)\\
$\bar{f}_u(f_u,f_n)=$  & $\left(f_n-1\right)^6 f_u \left(48 \left(f_n-1\right) f_u^6-168 \left(f_n-1\right) f_u^5+252 \left(f_n-1\right)
f_u^4-210 \left(f_n-1\right) f_u^3\right.$\\
& $\left.+14 \left(9 f_n-7\right) f_u^2+21 \left(1-3 f_n\right) f_u+21 f_n\right)$\\
$P_{\mathrm{succ}}(f_n)=$ & $\left(f_n-1\right)^6 \left(6 f_n+1\right)$ \\[1ex]
&7-qubit Steane code ($n_{\mathrm{max}}=2$)\\
$\bar{f}_u(f_u,f_n)=$  & $\left(f_n-1\right){}^5 f_u \left(48 \left(f_n-1\right){}^2 f_u^6-168 \left(f_n-1\right){}^2 f_u^5+252
\left(f_n-1\right){}^2 f_u^4-210 \left(f_n-1\right){}^2 f_u^3\right.$\\
&$\left.+14 \left(f_n \left(3 f_n-16\right)+7\right) f_u^2+21 \left(f_n \left(3 f_n+4\right)-1\right) f_u-21 f_n \left(2
f_n+1\right)\right)$ \\
$P_{\mathrm{succ}}(f_n)=$ & $-\left(f_n-1\right){}^5 \left(15 f_n^2+5 f_n+1\right)$\\[1ex]
&7-qubit Steane code ($n_{\mathrm{max}}=3$)\\
$\bar{f}_u(f_u,f_n)=$  & $\frac{1}{2} \left(f_n-1\right){}^4 \left(f_n^3 \left(96 f_u^7-336 f_u^6+504 f_u^5-420 f_u^4+308 f_u^3-210 f_u^2+84
f_u+7\right)\right.$\\
&$-2 f_n^2 f_u \left(144 f_u^6-504 f_u^5+756 f_u^4-630 f_u^3+266 f_u^2-21 f_u-21\right)$\\
&$+2 f_n f_u \left(144 f_u^6-504 f_u^5+756 f_u^4-630 f_u^3+322 f_u^2-105 f_u+21\right)$\\
&$\left.+2 f_u^2 \left(-48 f_u^5+168 f_u^4-252 f_u^3+210 f_u^2-98 f_u+21\right)\right)$\\
$P_{\mathrm{succ}}(f_n)=$ & $\left(f_n-1\right){}^4 \left(20 f_n^3+10 f_n^2+4 f_n+1\right)$\\[1ex]
&7-qubit Steane code ($n_{\mathrm{max}}=4$)\\
$\bar{f}_u(f_u,f_n)=$ & $\frac{1}{2} \left(f_n-1\right){}^3 \left(3 f_n^4 \left(32 f_u^7-112 f_u^6+168 f_u^5-140 f_u^4+84 f_u^3-42 f_u^2+14
f_u-7\right)\right.$\\
&$-f_n^3 \left(384 f_u^7-1344 f_u^6+2016 f_u^5-1680 f_u^4+840 f_u^3-252 f_u^2+42 f_u+7\right)$\\
&$+12 f_n^2 f_u^2 \left(48 f_u^5-168 f_u^4+252 f_u^3-210 f_u^2+98 f_u-21\right)$\\
&$-6 f_n f_u \left(64 f_u^6-224 f_u^5+336 f_u^4-280 f_u^3+140 f_u^2-42 f_u+7\right)$\\
&$\left.+2 f_u^2 \left(48 f_u^5-168 f_u^4+252 f_u^3-210 f_u^2+98 f_u-21\right)\right)$\\
$P_{\mathrm{succ}}(f_n)=$ & $-15 f_n^7+35 f_n^6-21 f_n^5+1$\\[1ex]
&7-qubit Steane code ($n_{\mathrm{max}}=5$)\\
$\bar{f}_u(f_u,f_n)=$ & $\frac{1}{2} \left(f_n-1\right){}^2 \left(6 f_n^5 f_u \left(16 f_u^6-56 f_u^5+84 f_u^4-70 f_u^3+42 f_u^2-21
f_u+7\right)\right.$\\
&$-2 f_n^4 \left(240 f_u^7-840 f_u^6+1260 f_u^5-1050 f_u^4+546 f_u^3-189 f_u^2+42 f_u-7\right)$\\
&$+f_n^3 \left(960 f_u^7-3360 f_u^6+5040 f_u^5-4200 f_u^4+2016 f_u^3-504 f_u^2+42 f_u+7\right)$\\
&$-6 f_n^2 f_u \left(160 f_u^6-560 f_u^5+840 f_u^4-700 f_u^3+336 f_u^2-84 f_u+7\right)$\\
&$+2 f_n f_u \left(240 f_u^6-840 f_u^5+1260 f_u^4-1050 f_u^3+518 f_u^2-147 f_u+21\right)$\\
&$\left.+2 f_u^2 \left(-48 f_u^5+168 f_u^4-252 f_u^3+210 f_u^2-98 f_u+21\right)\right)$\\
$P_{\mathrm{succ}}(f_n)=$ & $6 f_n^7-7 f_n^6+1$\\[1ex]
&7-qubit Steane code ($n_{\mathrm{max}}=6$)\\
$\bar{f}_u(f_u,f_n)=$ & $f_n^7 \left(48 f_u^7-168 f_u^6+252 f_u^5-210 f_u^4+126 f_u^3-63 f_u^2+21 f_u-\frac{7}{2}\right)$\\
&$-\frac{21}{2} f_n^6 \left(2 f_u-1\right){}^3 \left(4 f_u^4-8 f_u^3+6 f_u^2-2 f_u+1\right)$\\
&$+\frac{21}{2} f_n^5 \left(2 f_u-1\right){}^3 \left(12 f_u^4-24 f_u^3+18 f_u^2-6 f_u+1\right)$\\
&$-105 f_n^4 f_u \left(2 f_u-1\right){}^3 \left(2 f_u^3-4 f_u^2+3 f_u-1\right)+\frac{7}{2} f_n^3 \left(2 f_u-1\right){}^3 \left(60 f_u^4-120
f_u^3+90 f_u^2-30 f_u-1\right)$\\
&$-63 f_n^2 f_u \left(2 f_u-1\right){}^3 \left(2 f_u^3-4 f_u^2+3 f_u-1\right)+21 f_n f_u \left(2 f_u-1\right){}^3 \left(2 f_u^3-4 f_u^2+3
f_u-1\right)$\\
&$+f_u^2 \left(-48 f_u^5+168 f_u^4-252 f_u^3+210 f_u^2-98 f_u+21\right)$\\
$P_{\mathrm{succ}}(f_n)=$ & $1-f_n^7$\\[1ex]
&7-qubit Steane code ($n_{\mathrm{max}}=7$)\\
$\bar{f}_u(f_u,f_n)=$ & $3 f_n^7 \left(2 f_u-1\right){}^3 \left(2 f_u^4-4 f_u^3+3 f_u^2-f_u+1\right)-\frac{21}{2} f_n^6 \left(2
f_u-1\right){}^3 \left(4 f_u^4-8 f_u^3+6 f_u^2-2 f_u+1\right)$\\
&$+\frac{21}{2} f_n^5 \left(2 f_u-1\right){}^3 \left(12 f_u^4-24 f_u^3+18 f_u^2-6 f_u+1\right)-105 f_n^4 f_u \left(2 f_u-1\right){}^3
\left(2 f_u^3-4 f_u^2+3 f_u-1\right)$\\
&$+\frac{7}{2} f_n^3 \left(2 f_u-1\right){}^3 \left(60 f_u^4-120 f_u^3+90 f_u^2-30 f_u-1\right)-63 f_n^2 f_u \left(2 f_u-1\right){}^3
\left(2 f_u^3-4 f_u^2+3 f_u-1\right)$\\
&$+21 f_n f_u \left(2 f_u-1\right){}^3 \left(2 f_u^3-4 f_u^2+3 f_u-1\right)+f_u^2 \left(-48 f_u^5+168 f_u^4-252 f_u^3+210 f_u^2-98
f_u+21\right)$\\
$P_{\mathrm{succ}}(f_n)=$ & $1$\\
\end{tabular}
\end{table*}
\begin{table*}[hpt]
\caption{Logical error rate for the Golay-Code}
\begin{tabular}{rl}
& Golay code\cite{Elia95} assuming $\bar{f}_u\approx \frac{p_w}{2}$\\
$\bar{f}_u(f_u,f_n)=$ & $\frac{1}{2} \left(-\frac{f_n^{23}}{4096}+\frac{23 \left(f_n+f_u-1\right) f_n^{22}}{2048}-\frac{253
\left(f_n+f_u-1\right){}^2 f_n^{21}}{1024}+\frac{1771}{512} \left(f_n+f_u-1\right){}^3 f_n^{20}-\frac{8855}{256} \left(f_n+f_u-1\right){}^4
f_n^{19}\right.$\\
&$+\frac{33649}{128} \left(f_n+f_u-1\right){}^5 f_n^{18}-\frac{100947}{64} \left(f_n+f_u-1\right){}^6 f_n^{17}+\frac{245157}{32}
\left(f_n+f_u-1\right){}^7 f_n^{16}-30613 \left(f_n+f_u-1\right){}^8 f_n^{15}$\\
&$-\frac{253}{16} \left(f_n-1\right) \left(f_n+f_u-1\right){}^7 f_n^{15}+101200 \left(f_n+f_u-1\right){}^9 f_n^{14}$\\
&$+\frac{3795}{8} \left(f_n-1\right) \left(f_n+f_u-1\right){}^8 f_n^{14}-272734 \left(f_n+f_u-1\right){}^{10} f_n^{13}-\frac{26565}{4}
\left(f_n-1\right) \left(f_n+f_u-1\right){}^9 f_n^{13}$\\
&$+560924 \left(f_n+f_u-1\right){}^{11} f_n^{12}+\frac{115115}{2} \left(f_n-1\right) \left(f_n+f_u-1\right){}^{10} f_n^{12}-695520
\left(f_n+f_u-1\right){}^{12} f_n^{11}$\\
&$-319424 \left(f_n-1\right) \left(f_n+f_u-1\right){}^{11} f_n^{11}+\frac{8855}{2} \left(f_n+f_u-1\right){}^{11} \left(-f_n+2 f_u+1\right)
f_n^{11}$\\
&$+949256 \left(f_n-1\right) \left(f_n+f_u-1\right){}^{12} f_n^{10}-97405 \left(f_n+f_u-1\right){}^{12} \left(-f_n+2 f_u+1\right)
f_n^{10}$\\
&$+779240 \left(f_n+f_u-1\right){}^{13} \left(-f_n+2 f_u+1\right) f_n^9+18975 \left(f_n+f_u-1\right){}^{13} \left(-f_n+6 f_u+1\right)
f_n^9$\\
&$-485760 \left(f_n+f_u-1\right){}^{14} \left(-f_n+6 f_u+1\right) f_n^8-2277 \left(f_n+f_u-1\right){}^{14} \left(-f_n+14 f_u+1\right)
f_n^8$\\
&$+32384 \left(f_n+f_u-1\right){}^{15} \left(-f_n+14 f_u+1\right) f_n^7+\frac{253}{2} \left(f_n-1\right) \left(f_n+f_u-1\right){}^{14}
\left(-f_n+14 f_u+1\right) f_n^7$\\
&$+212520 \left(f_n+f_u-1\right){}^{14} \left(-\left(f_n-1\right){}^2+10 f_u \left(f_n-1\right)+8 f_u^2\right) f_n^7$\\
&$-100947 \left(f_n-1\right) \left(f_n+f_u-1\right){}^{15} \left(-f_n+14 f_u+1\right) f_n^6$\\
&$-28336 \left(f_n+f_u-1\right){}^{16} \left(-f_n+2 f_u+1\right) \left(-f_n+14 f_u+1\right) f_n^5$\\
&$-5313 \left(f_n+f_u-1\right){}^{16} \left(\left(f_n-1\right){}^2-15 f_u \left(f_n-1\right)+30 f_u^2\right) f_n^5$\\
&$+8855 \left(f_n+f_u-1\right){}^{17} \left(\left(f_n-1\right){}^2-17 f_u \left(f_n-1\right)+90 f_u^2\right) f_n^4$\\
&$-1771 \left(f_n+f_u-1\right){}^{17} \left(\left(f_n-1\right){}^3-17 f_u \left(f_n-1\right){}^2+138 f_u^2 \left(f_n-1\right)+96
f_u^3\right) f_n^3$\\
&$-253 \left(f_n+f_u-1\right){}^{18} \left(-\left(f_n-1\right){}^3+18 f_u \left(f_n-1\right){}^2-171 f_u^2 \left(f_n-1\right)+90
f_u^3\right) f_n^2$\\
&$+23 \left(f_n+f_u-1\right){}^{19} \left(-\left(f_n-1\right){}^3+19 f_u \left(f_n-1\right){}^2-190 f_u^2 \left(f_n-1\right)+560
f_u^3\right) f_n+\left(f_n+f_u-1\right){}^{23}$\\
&$\left.-23 f_u \left(f_n+f_u-1\right){}^{22}+253 f_u^2 \left(f_n+f_u-1\right){}^{21}-1771 f_u^3
\left(f_n+f_u-1\right){}^{20}+1\vphantom{-\frac{f_n^{23}}{4096}+\frac{23 \left(f_n+f_u-1\right) f_n^{22}}{2048}-\frac{253
\left(f_n+f_u-1\right){}^2 f_n^{21}}{1024}+\frac{1771}{512} \left(f_n+f_u-1\right){}^3 f_n^{20}-\frac{8855}{256} \left(f_n+f_u-1\right){}^4
f_n^{19}}\right)$\\
$P_{\mathrm{succ}}=$ & $1$
\end{tabular}
\end{table*}

\end{document}